  \documentclass{aa} 

  \def\swift{{\it Gehrels Swift~}} 
   
  \def\nustar{{\it NuSTAR~}} 
  \def\funits{$\rm erg\,cm^{-2}\,s^{-1}$~}
  \def\lunits{$\rm erg\,s^{-1}$~}

  \usepackage{natbib}
  \usepackage{graphicx}
  \usepackage{longtable}

  \usepackage{txfonts}

\begin{document}

     \title{The distribution of the coronal temperature in Seyfert 1 galaxies}

     
      \titlerunning{The distribution of the coronal temperature in Seyfert 1 galaxies}
      \authorrunning{A. Akylas et al.}
     
     \author{A. Akylas
            \inst{1}
            \and
            I. Georgantopoulos
            \inst{1}
                        }

     \institute{Institute for Astronomy Astrophysics Space Applications and Remote Sensing (IAASARS), National Observatory of Athens, I. Metaxa \& V. Pavlou, Penteli, 15236, Greece\\
                \email{aakylas@noa.gr}
                      }


    \abstract{
    The Active Galactic Nuclei (AGN) produce copious amounts of X-rays through 
    the corona that is the hot gas that lies close to the accretion disk. 
    The temperature of the corona can be accurately determined by the cut-off signature in the X-ray spectrum.  Owing to the large temperatures of the corona, observations well above 10 keV are necessary.
    Here, we explore the  {\it NuSTAR} observations of 
    118 {\it Gehrels/Swift} selected Seyfert 1 AGN. We model the spectrum using a single power-law with 
    an exponential cut-off modified by neutral and ionised absorption as well as a reflection component. We find secure spectral cut-off estimates in 62 sources while for the remaining ones we derive only lower limits.
     The mean value is 103 keV with a skewed distribution towards large energies with large dispersion. When we consider the lower limits using survival analysis techniques, the mean cut-off energy becomes significantly larger,  about 200 keV.  Because of various limitations (e.g. limited spectral passband, photon statistics, model degeneracies)  we perform extensive simulations to explore the underlying  spectral cut-off distribution. We find that an intrinsic spectral cut-off distribution which has a Maxwell-Boltzmann shape with a mean value in the range of 160 - 200 keV can reproduce sufficiently well the observations. Finally, our spectral analysis places very stringent constraints on both the photon index ($\Gamma=1.77\pm0.01$) as well as on the  reflection component ($\rm R=0.69\pm0.04$) of the Seyfert 1 population.  From the values of the spectral cut-off and the photon-index we deduce that the mean optical depth of the AGN corona is approximately $\rm \tau_e=1.82\pm0.14$ and its mean temperature approximately $\rm kT_e=65\pm10$ keV. 
     }
    

    \authorrunning{Akylas et al.}

    \maketitle
  %

	\section{Introduction}
	Accreting SMBH (Supermassive Black Holes) or AGN (Active Galactic Nuclei) produce large amounts of radiation primarily in the  UV and optical wavelengths.
	This radiation is  believed to be produced in a optically thick accretion disk formed by material infalling to the black hole. 
	X-ray radiation is a ubiquitous feature of AGN \citep{brandt2015}. 
	The X-ray continuum is produced in a region of hot plasma called the corona. 
	The X-rays are produced  as the hot electrons scatter the UV photons coming from the accretion disk through inverse Compton scattering \citep{vaiana1978,haardt1991}. The rapid X-ray variability 
	observed \citep[e.g.][]{mchardy2005,ludlam2015}, as well as reverberation of the X-ray radiation reprocessed by the accretion disk, the so-called reflection component,  \citep{emmanoulopoulos2014,uttley2014, kara2016} suggest that the corona is small in size (several times the gravitational radius of the black hole). These findings have been confirmed by microlensing studies \citep[e.g.][]{chartas2016, guerras2017}.
	
	Broad-band X-ray spectroscopy of the X-ray emission 
	can provide important constraints on the physical parameters of the
	coronal gas. In particular, the temperature $\rm kT_e$ in combination with the 
	optical depth $\tau_e$ determines the slope of the power-law spectrum 
	\citep[e.g][]{petrucci2001}. The higher the optical depth, the flatter the 
	X-ray continuum at a given temperature of the corona. The X-ray broad-band spectrum 
	is usually parameterised as $E^{-\Gamma}e^{-E/E_c}$ where E is the photon energy, 
	$\Gamma$ the photon index and $E_c$ is the spectral energy cut-off that accurately measures the temperature of the corona.  However, it is 
	challenging to determine the value of the the spectral cut-off as this 
	requires observations well above 10 keV. 
	The spectral cut-off has been measured directly for the first time using the
	{\it Compton Gamma-ray} observatory. It has been found that the spectral cut-off 
	in NGC4151 is around 100 keV \citep{johnson1997}. 
	The {\it BeppoSAX} mission increased the number of the spectral cut-off measurements 
	\citep{dadina2007}.
	More recently, a large number of observations has been accumulated using either 
	the {\it INTEGRAL} or the {\it Gehrels SWIFT} missions. 
	\citet{malizia2014} presented the {\it INTEGRAL} spectra of a sample of Seyfert 1 galaxies detecting securely the cut-off energy in 26 sources. They find an average high energy cut-off of 128 keV with a standard deviation of 46 keV. \citet{ricci2017} presented the most comprehensive X-ray spectral analysis of AGN in the local Universe. They presented the X-ray spectra of all the AGN detected in the {\it Gehrels SWIFT}/BAT survey. 
	There are 352 unobscured AGN ($\rm N_H<10^{22} cm^{-2})$. In 
	89 sources the spectral cut-off could be securely inferred (no censored values, i.e. upper or lower limits) with a mean value of $80\pm7$ keV. Taking into account
	all sources i.e. including lower and upper limits by means of survival analysis techniques, they find a mean cut-off energy of $331\pm29$ keV. 
	This  value is substantially higher than that derived by \citet{malizia2014}. 
	
	The launch of the {\it NuSTAR} mission \citep{harrison2013} brought a leap forward  in the study of the high energy spectra of AGN owing to its excellent spatial resolution
	 above 10keV.  {\it NuSTAR} observations of the cut-off of AGN include  \citet{fabian2015}, \citet{kamraj2018}, \citet{molina2019}, \citet{rani2019}. 
	 \citet{molina2019} analysing a sample of 18 Seyfert 1 galaxies, 13 secure  measurements plus five lower limits, find a a mean value for the cut-off of  $\rm E_c=111~ keV$ with a dispersion of 45 keV. \citet{rani2019} present a sample of 
	  eight Seyfert 1 AGN with securely derived cut-off.  This sample yields a mean cut-off value of 95 keV with a  dispersion of 32 keV. 
	 Finally, \citet{balokovic2020} presented the {\it NuSTAR} spectra of 130  Seyfert 2 galaxies selected by {\it Gehrels SWIFT}. They find a median  cut-off of $290\pm20$ keV. 
	 
	 It becomes evident that there are still significant uncertainties in the value of 
	 the spectral cut-off and hence the temperature of the corona. This is partly because 
	 of the presence of a considerable fraction of lower limits among some of the above samples. Moreover, there is a considerable spread in the values of the cut-offs even among the securely constrained values. It is not clear whether this large spread  reflects the intrinsic dispersion of AGN coronal temperatures or alternatively  
	 it can be attributed to the limited photon statistics and 
	 the relatively  limited pass-band of even the hard-energy detectors.
	 In this paper we are trying to overcome this impasse by exploring the X-ray spectroscopic analysis, with {\it NuSTAR}, of the largest sample of Seyfert 1 galaxies presented so far. The Seyfert 1 sample comes from the 107-month {\it  Gehrels SWIFT}/BAT sample of \citet{oh2018}.  Our primary goal is to derive the intrinsic X-ray spectral cut-off distribution. 
	 A key feature of our 
	 analysis is that we present detailed simulations in order to  take into account all systematics that affect the  determination of the spectral cut-off.
	 Together with the robust estimation of the spectral cut-off, our 
	 spectral modelling provides strong constraints on the photon index distribution as well as on the strength of the reflection component. 
	 Throughout the paper, we adopt the standard cosmological parameters $\rm H_o=70km s^{-1}Mpc^{-1}$, $\rm \Omega_m=0.3, \Omega_\Lambda=0.7$. 

  \section{The sample}\label{thesample}
  
    Here we compile the archival Nuclear Spectroscopic Telescope Array (\nustar) observations of the Seyfert 1 
    galaxies detected in the 105-month survey of the Burst Alert
   	Telescope, BAT, \citep{Barthelmy2004} survey on-board the \swift Gamma-Ray Burst observatory \citep{gehrels2004}.
   	The 105-month BAT survey \citep{oh2018} is a uniform, hard X-ray, all-sky survey with a sensitivity of
   	8.40$\rm \times10^{-12} erg~s^{-1}$ cm$^{-2}$ over 90\% of the sky and 7.24$\rm \times10^{-12} \rm erg~s^{-1}cm^{-2}$ over 50\% of the sky, in the $\rm14-195$ keV band. 
   	The BAT 105 month catalogue provides 1632 hard X-ray sources in the $14-195$ keV band above the 4.8$\sigma$ significance level.
   	Frequently,  since its start of operations in 2012, 
   	the {\it NuSTAR} satellite has been taking  observations of AGNs selected from the 
   	{\it Gehrels Swift}/BAT hard  X-ray catalogue.
 
  	\nustar,  \citep{harrison2013} launched in June 2012, is the first orbiting X-ray observatory which
   	focuses light at high energies (E $>$ 10 keV). It consists of two
   	co-aligned focal plane modules (FPMs), which are identical in design. Each
   	FPM covers the same 12 x 12 arcmin portion of the sky, and comprises of
   	four Cadmium-Zinc-Tellurium detectors.  \nustar operates between 3 and 79 
   	keV and provides an improvement of at least two orders of magnitude  in 
   	sensitivity compared to previous hard X-ray observatories operating at 
   	energies E$>$10 keV.  We take advantage of the \nustar unprecedented sensitivity
   	above 10 keV to measure, the distribution  of the
   	high energy cut-off of the Seyfert 1 population in the local universe . 

   	In the 105 months {\it Gehrels Swift}/BAT hard  X-ray catalogue there are in total 370
   	Seyfert 1 galaxies 
   	including type 1.0 (163 AGN) type 1.2 (96 AGN) and type 1.5 (111 AGN). Up until recently, there 
   	have been observed in total 142 of these {\it Gehrels Swift/BAT} Seyfert 1  with {\it NuSTAR}.
   	Note that the majority of the sources observed by {\it NuSTAR} belong to the 70-month BAT catalogue 
   	\citep{baumgartner2013} while only an additional four come from the 105 month catalogue.
   	The effective energy range of \nustar detectors (3-79 keV)  does not allow an 
   	accurate measurement of the lower $\rm N_H$ hydrogen  column densities.  Despite the fact that \nustar is  not 
   	sensitive to soft X-rays below 3 keV, even moderate column densities could easily absorb the X-ray photons within its pass-band.
   	For example assuming a power-law spectrum with a photon index $\Gamma=1.8$, 
   	a column density of $\rm N_H=10^{22} cm^{-2}$ reduces the \nustar flux in the 3-10 keV band by
   	$\sim$4 per cent. If the column density $\rm N_H$ becomes $\rm 5\times 10^{22} cm^{-2}$ then the reduction of 
   	the 3-10 keV flux  becomes $\sim18$ per cent. 
   	In order to minimize the effect of the photo-electric
   	absorption at the lower part of the X-ray spectrum  we focus only on the X-ray unobscured Seyfert 1  galaxies 
   	(i.e. $\rm N_H$ <$\rm 1\times 10^{22} cm^{-2}$). 
   	This is important since the degeneracy between the photon index and the column density may affect the cut-off
   	energy measurements.
   	
   	\cite{ricci2017} have provided accurate measurements of the X-ray absorption by using {\it XMM-Newton}, {\it 
   	Swift}/XRT, {\it ASCA}, {\it Chandra}, and {\it Suzaku} observations in the soft X-ray band ($<$10keV) with the 70-month averaged {\it Swift}/BAT data. Based on their results we  exclude from the analysis 20 sources showing  $\rm N_H >1 \times 10^{22} cm^{-2}$. Our final sample comprises of 122 sources.  The	observational details of the sample are listed in Table
   	\ref{obs_log} in increasing BAT obsID order. The source name, redshift  and the optical
   	classification type has been obtained  from the 105-month BAT survey catalogue. The \nustar
   	observation identifier used in the analysis is also shown. When multiple observations are present we choose to analyze the observation with the highest exposure. Last, the net counts of each observation calculated from the combined (FMPA and FMPB) observations are also presented.   
 
    \begin{figure}
    \begin{center}
    \includegraphics[height=0.75\columnwidth]{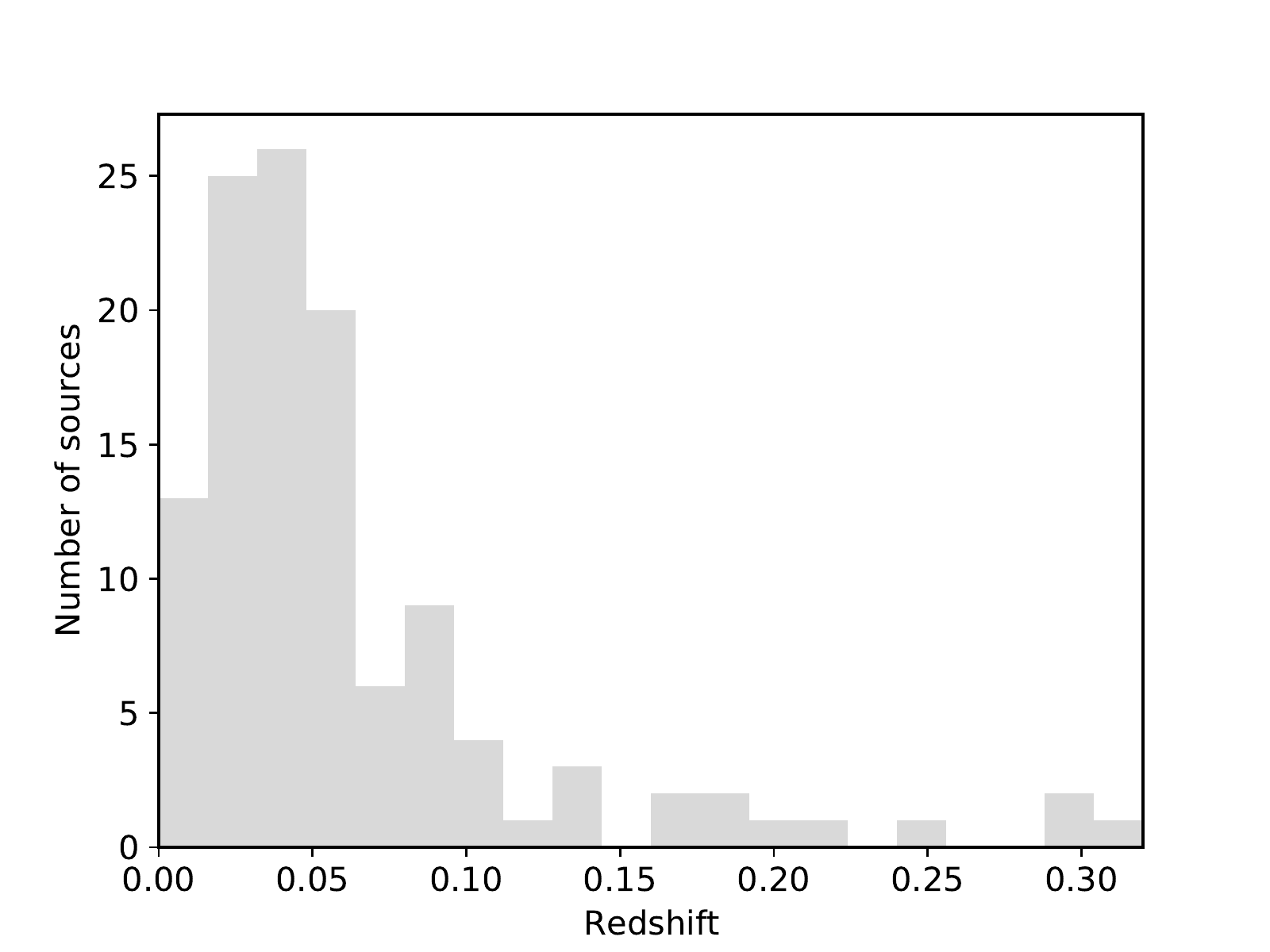}
    \end{center}
    \caption{The redshift distribution of the 122 sources in the final sample.}
    \label{zdist}
    \end{figure} 
	
	\section{X-ray spectral analysis}
	
	\subsection{X-ray spectral reduction}

    Spectral reduction has been performed for both {\it NuSTAR} modules, FPMA, and FPMB, using the {\it NuSTAR} Data Analysis Software ({\sc NuSTARDAS}; version 1.2.1), within the {\sc HEASOFT} (version 6.16). We have extracted source and background energy spectra from the calibrated and cleaned event files using the {\sc nuproducts} module. Detailed information on the data reduction procedures can be found in the NuSTAR Data Analysis Software Guide \citep{perri+2017}. An extraction radius of 60'' has been used for both the source and background regions. The background spectrum has been estimated from several source-free regions of the image at an off-axis angle similar  to the source position. The spectral files were re-binned using the {\sc HEASOFT} task {\sc GRPPHA} to give a minimum of 20 photon counts per bin. 
    In  Fig. \ref{obs_log} we present the background subtracted count distribution of our observations from both FPMA and FPMB instruments to depict the excellent statistical quality of the spectra. The minimum number of the background subtracted (FMPA \& FMPB) counts is 550  and corresponds to the source with BAT obsID 495 while BAT source 694 show the maximum number of 697782 counts. 

    \begin{figure}
    \begin{center}
    \includegraphics[height=0.75\columnwidth]{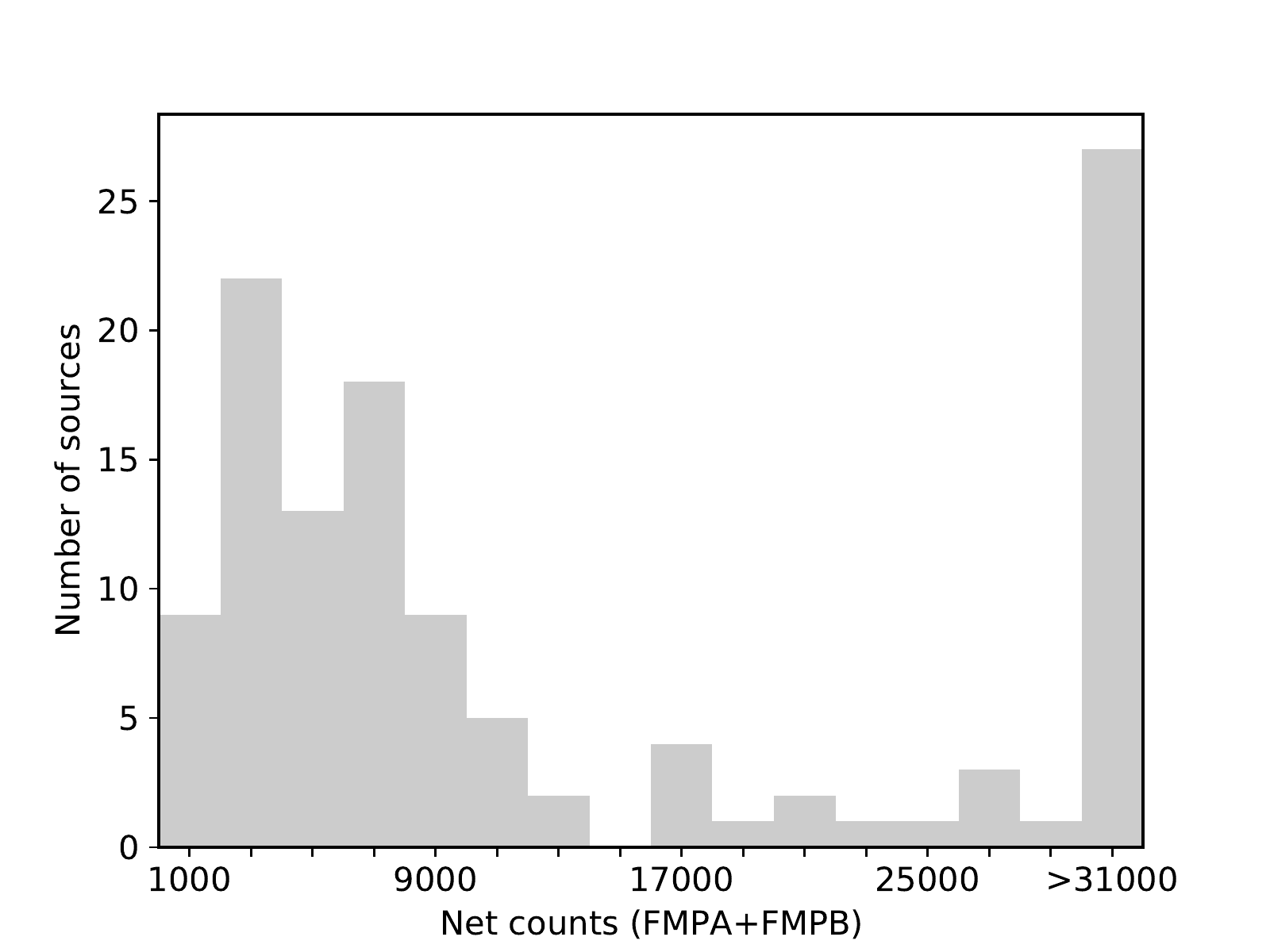}
    \end{center}
    \caption{The background subtracted count distribution of our observations from the combined FPMA and FPMB instruments. The minimum number of counts is 550 and the maximum number is 697782 counts. }
    \label{count_dist}
    \end{figure}
    
    \subsection{Spectral fitting}{\label{fitting}}
    
    The spectral fitting is carried out using {\sc XSPEC} v12.8.2 \citep{arnaud1996}. We simultaneously fit the spectra from both the FMPA and FMPB instruments.   We fit the data using the {\sc PEXMON} model in {\sc XSPEC} \citep{Nandra2007}. This describes an exponentially cut-off power-law spectrum, reflected by a neutral material slab and takes into account the self-consistently generated narrow Fe K lines. The relative reflection strength is parameterised by R, where R=1 corresponds to a semi-infinite slab seen at an inclination angle of $\theta$ and subtends a $2\pi$ solid angle at the X-ray source. We take both neutral and ionized absorption into account. Although we have already discarded all sources with absorption above $\rm 10^{22} cm^{-2}$ as found by \citet{ricci2017}, we check for possible variability in the column densities. The neutral absorption column density is modeled using the {\sc ZPHABS} model in {\sc XSPEC}. Regarding ionised absorption, the hard spectral band of {\it NuSTAR} which excludes energies below 3 keV, does not allow us to easily constrain any possible ionised absorption features. Using softer X-ray data, \citet{ricci2017} find evidence for ionised absorption in 33 sources from our sample. This was found using the {\sc ZXIPCF} model \citep{reeves2008} which uses a grid of {\sc XSTAR} absorption models \citep{kallman2001}. We fix the setting of  the ionised absorption values to those derived by \citet{ricci2017}. 
     Following \citet{ricci2017} and since we are primarily interested on the spectral cut-off, we chose for simplicity to model the Fe line emission only using a narrow component. Then, our modelling of the Fe emission relied on the {\sc PEXMON} model and we do not add a broad Fe line component. Finally, a constant multiplication factor, varying within 3 per cent between FMP instruments, has also been  included to account for calibration issues.
    In {\sc XSPEC} notation our model is described as:  { \sc ZPHABS*ZXIPCF*PEXMON }.
    The inclination angle of the reflecting slab has been frozen to 60 degrees.  The photon index ($\rm \Gamma$), the relative reflection strength (R), the high energy cutoff ($\rm E_c$) and the normalization parameters are free to vary. We use $\chi^2$ statistics for  goodness of fitting and error estimation. 

\begin{figure}
    \begin{center}
    \includegraphics[height=0.75\columnwidth]{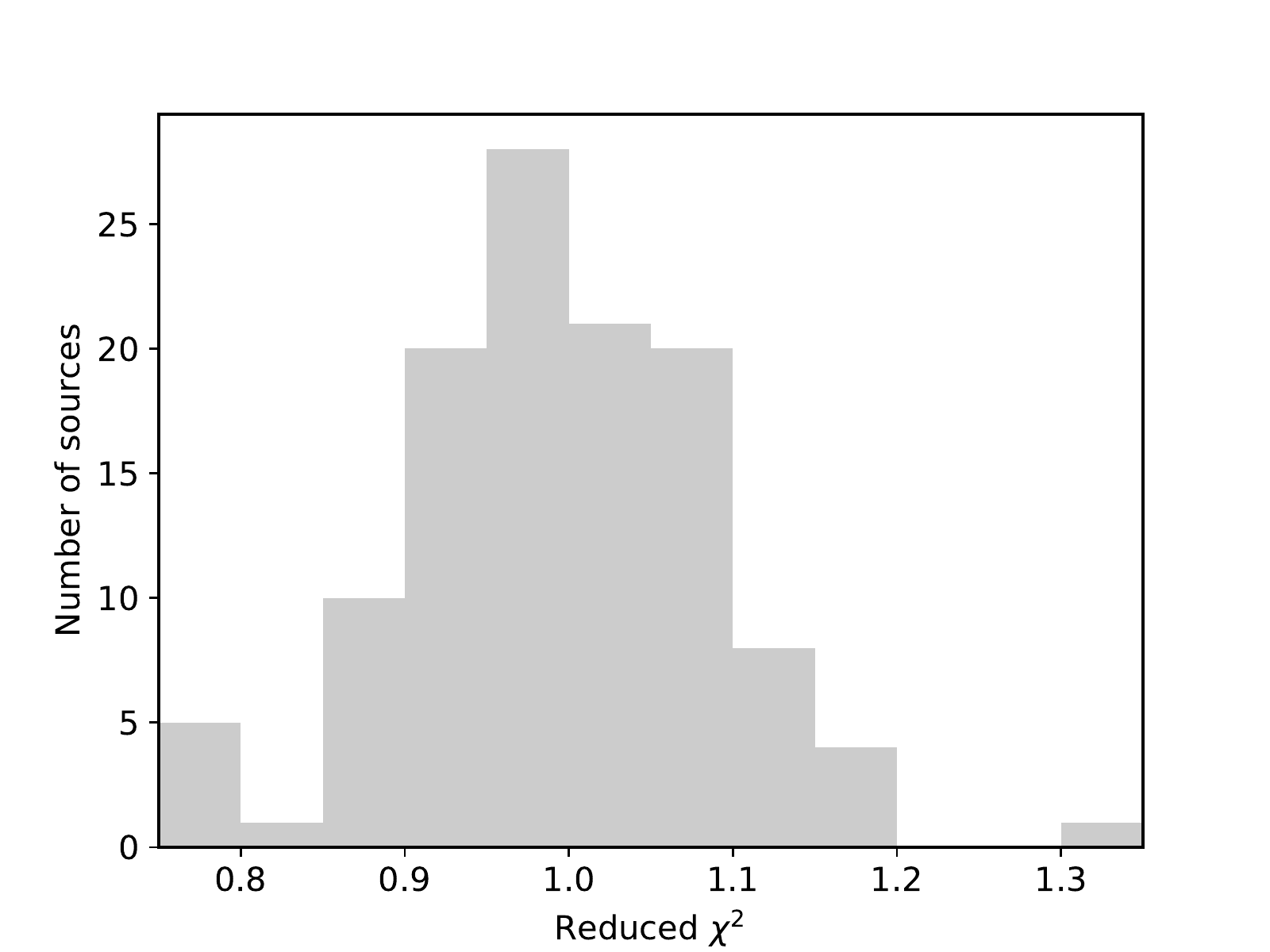}
    \end{center}
    \caption{The reduced $\chi^2$ ($\chi^2$ over degrees of freedom) for all sources.}
    \label{chi_dist}
    \end{figure}

\begin{figure}
    \begin{center}
    \includegraphics[height=0.75\columnwidth]{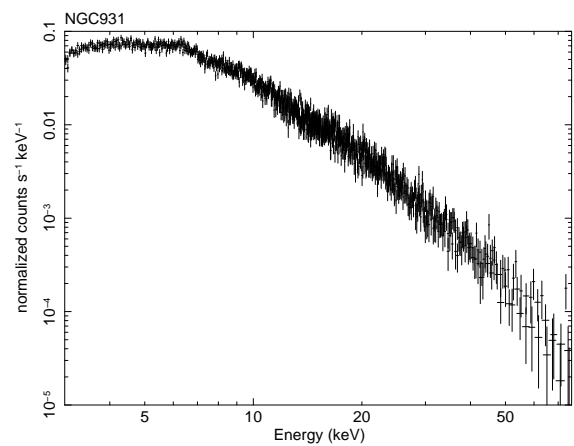}
    \end{center}
    \caption{The spectrum of the source NGC931.}
    \label{example_spec}
    \end{figure}

    In Fig. \ref{chi_dist} we show the distribution of the reduced  $\chi^2$ ($\chi^2$ over degrees of freedom) for all our spectra. In order to demonstrate the excellent quality of the photon statistics we present as an example the spectrum of the source NGC931
    in Fig. \ref{example_spec}.
   
    \section{Results}
    
    In Table \ref{spectral_fit} we present all the spectral fitting results for each source and the corresponding BAT ObsID, in increasing order, to facilitate direct comparison with the information listed in Table \ref{obs_log}.  All the errors in the spectral components correspond to the 90 per cent confidence interval.  The estimated flux and luminosity are also presented for both the soft (2-10 keV) and the hard (20-40 keV) bands. 
    
    \subsection{Absorption}
    Although we omitted from our sample all sources with column density 
    $\rm N_H>10^{22}~cm^{-2}$ as estimated in \citet{ricci2017}, our spectral fitting 
     revealed four sources that present significant absorption ($\rm >5\times10^{22} cm^{-2}$) in the {\it NuSTAR} spectra. These are the sources with BAT obsID 449, 765, 912 and 976. As all these appear unobscured in \citet{ricci2017} analysis, this suggests significant variability in the obscuring screen. At least for the case of Mrk704, such evidence have been presented previously in e.g. \citet{matt2011}. Following our selection criteria, discussed in Section \ref{thesample} we exclude these four sources from further analysis, leaving 118 sources in our sample.
    
    \subsection{Photon index}
    
\begin{figure}
    \begin{center}
    \includegraphics[height=0.75\columnwidth]{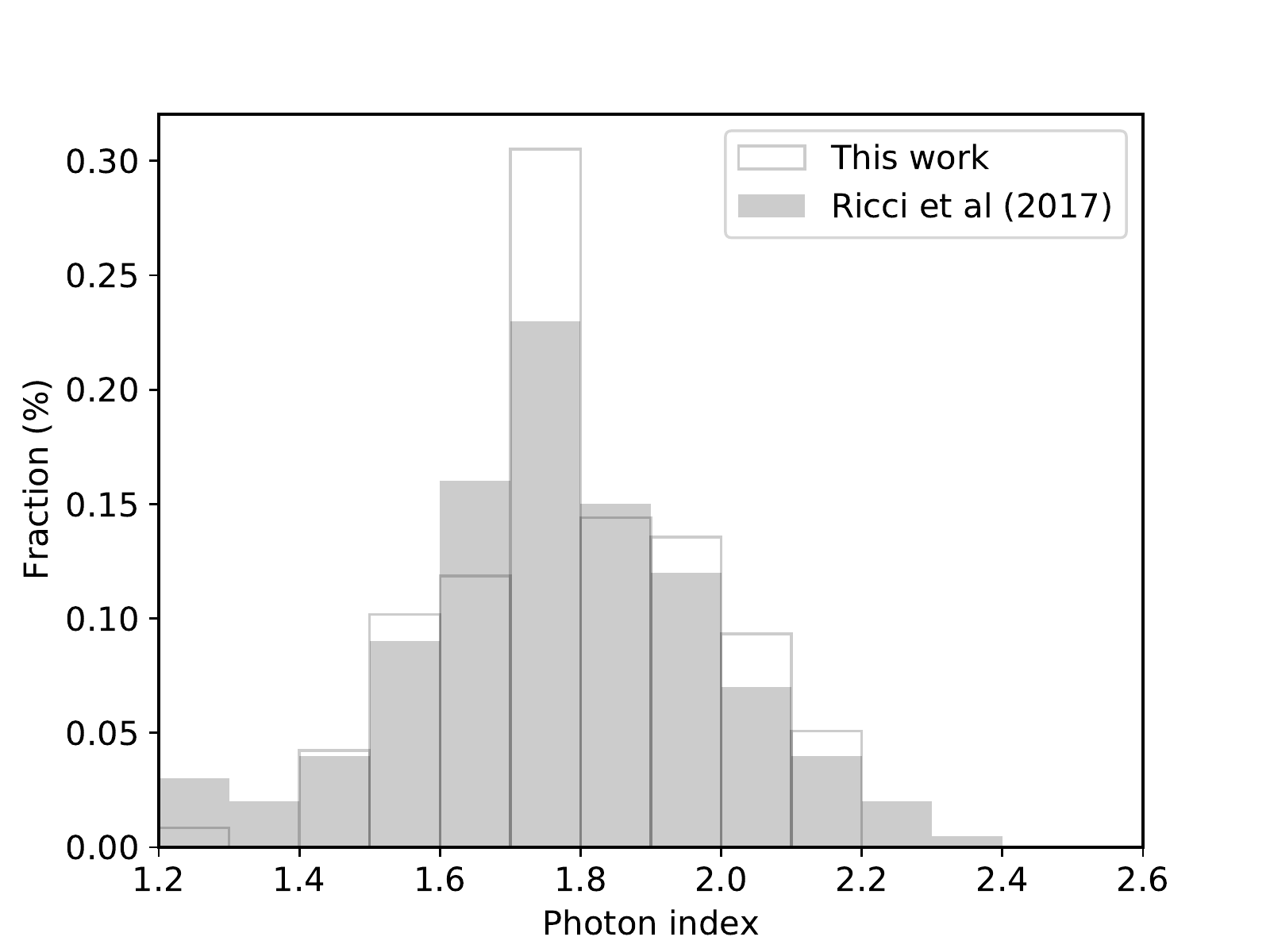}
    \end{center}
    \caption{Distribution of the photon indices (open histogram). The gray histogram shows the estimates from \citet{ricci2017}. The distributions are normalized to facilitate the comparison.}
    \label{g_dist}
    \end{figure}

     The photon index has been accurately measured for all  118 sources with no upper or lower limit measurements. The normalized distribution is presented in Fig. \ref{g_dist}. The median (mean) value of the photon index distribution is 1.77 (1.78). The upper and lower quartiles are 0.13 and 0.14 respectively. We can  approximate the 90 per cent error of the median (mean) using the relations $\sigma_{mean}=\sqrt{(\sum (\delta\Gamma_{i})^2)}/N$ and $\rm \sigma_{median}=k\times\sigma_{mean}$. The scaling factor k equals 1.253 assuming a Gaussian distribution and $\delta\Gamma_{i}$, the individual errors in the photon index, can be approximated by the average value of upper and lower limit uncertainties. Then the corresponding median (mean) error is $\sigma$=0.013 (0.010).
     
     \citet{ricci2017} presented the broadband X-ray characteristics of the 70-month {\it Swift}/BAT all-sky survey by combining BAT AGN spectra with deeper soft X-ray observations. Their best-fitting photon indices  for non-blazar AGNs with $\rm N_H<10^{22} cm^{-2}$  have also  been plotted for comparison in Fig. \ref{g_dist}. They find a median $\Gamma=1.80\pm0.02$. Both the  median value and the full distribution are in excellent agreement with our findings.

    \subsection{Reflection parameter}

\begin{figure}
    \begin{center}
    \includegraphics[height=0.75\columnwidth]{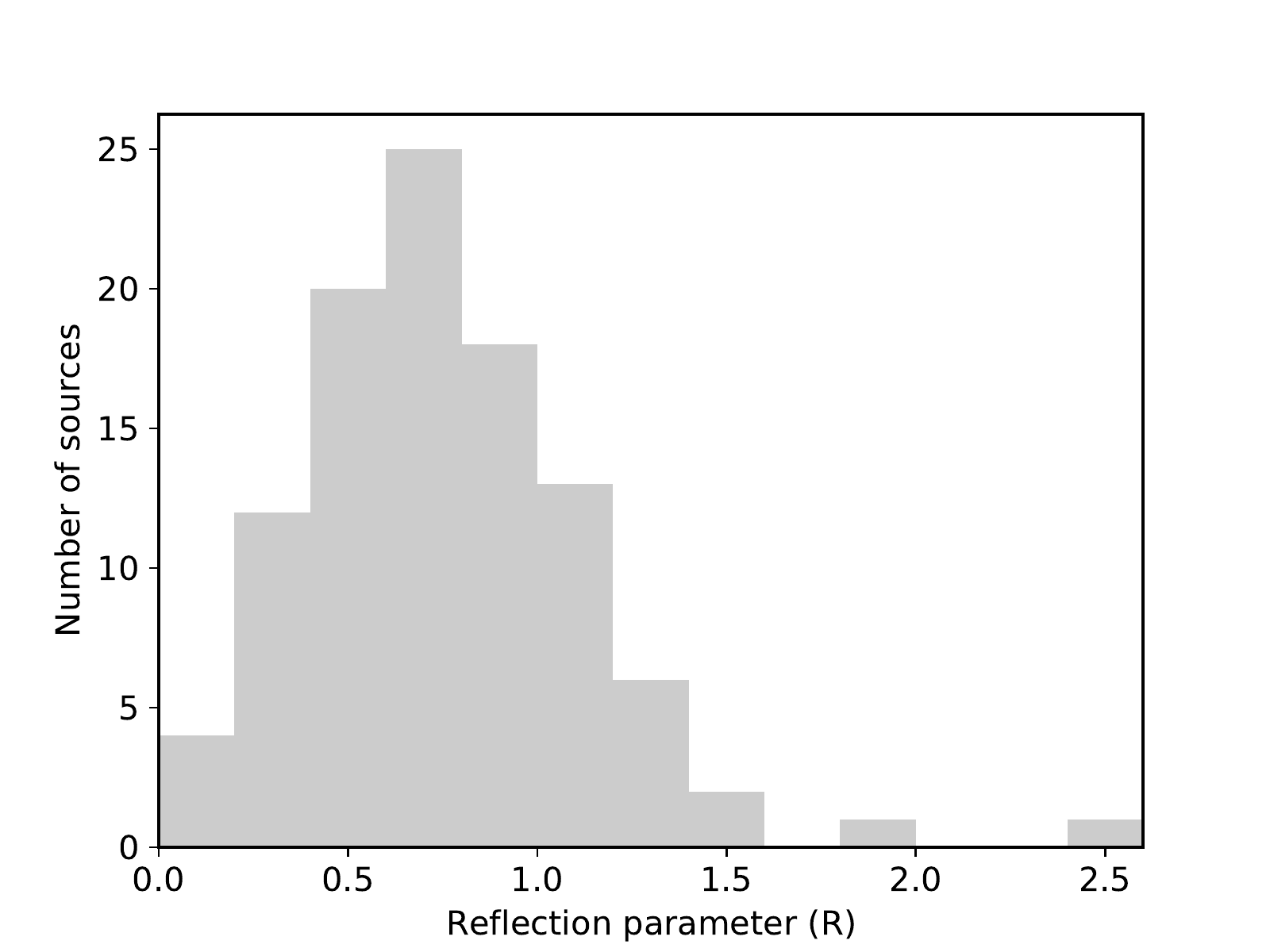}
    \end{center}
    \caption{Distribution of the reflection parameter R for the 106 sources in our sample.}
    \label{r_dist}
    \end{figure}

    The reflection parameter R, has been  securely inferred (no censored values)  for 106 out of 118 sources. For  12 sources ($\sim$ 13 per cent of the sample) only upper limits could be obtained for R.   In these cases the R value in Table \ref{spectral_fit} corresponds to the 90 per cent upper limit. 
    In Fig. \ref{r_dist} we plot the distribution of the reflection  parameter for the 106 sources. The median (mean) value is 0.71 (0.66). The first and third quartiles are 0.49 and 0.91 respectively. Using the same assumptions as in the case of the photon index the error on the median (mean) value is $\sigma$=0.06 (0.05).
     Next, we derive the reflection parameter, taking all 118 sources into account i.e. including the lower limits. We use the survival analysis {\sc ASURV} software package \citep{isobe1986}. 

     The derived corrected mean value from {\sc ASURV} is $R=0.69\pm0.04$.
     \citet{ricci2017} obtain a median R value of 0.83$\pm0.14$ taking into account all measurements for their $\rm N_H<10^{22} cm^{-2}$ sample. 
     This is in reasonable agreement with our estimates. However, the present {\it NuSTAR} results offer a substantial qualitative improvement over the previous BAT results. 
     This is because there is a significant fraction of lower limits in \citet{ricci2017}  while in our sample it is very low (13 per cent). 

    \subsection{The high energy cut-off}

    The high energy cut-off parameter, $\rm E_c$, has been securely measured for 62 sources (53 per cent of sample).  
    For the remaining 56 sources we provide the 90 per cent lower limit. 
    The highest $\rm E_c$ value of $334^{+646}_{-140}$keV measured in NGC1566, demonstrates the ability of  {\it NuSTAR} to constrain even  some highest energy cut-offs.  
    NGC4051 possibly represents another case of a cut-off at very high energies. Despite the lower-limit $E_c$ estimate for NGC4051, the results suggest a high energy cut-off greater than $\sim$700 keV at a 90 per cent confidence. Given the excellent quality of the spectrum, presenting more than 200,000 net counts, this suggests an extremely high value for $\rm E_c$ similar for example to the Seyfert 2 NGC5506 \citep{matt2015}. 
    
\begin{figure}
    \begin{center}
    \includegraphics[height=0.75\columnwidth]{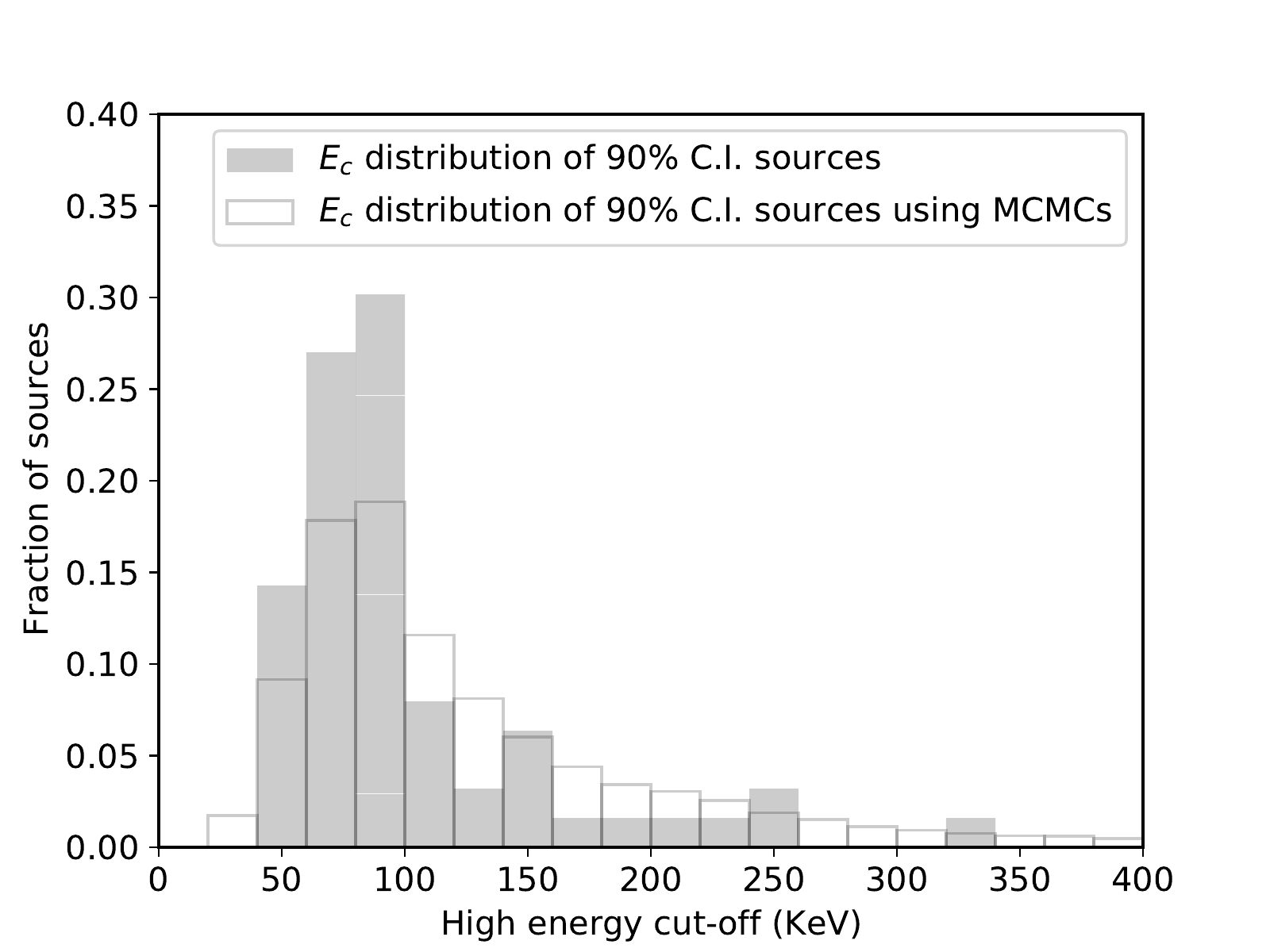}
    \end{center}
    \caption{Grey histogram: The observed distribution of the high energy spectral cut-off for the 62 sources  with securely inferred values. White histogram: The same population with the grey histogram after taking into account the uncertainties in each individual measurement of the $\rm E_c$ parameter using MCMC simulations.}
    \label{ec_dist}
    \end{figure}

    In Fig. \ref{ec_dist} we plot the distribution of the high energy cut-off  for the 62 sources with securely inferred spectral cut-offs. The median (mean) value is 89 (102) keV. The first and third quartiles are 65 keV and 102 keV respectively.  
    Using the same assumptions as in the case of the photon index and R the approximated error on the median (mean) value is $\sigma$=20 (16) keV.
    We note that \citet{ricci2017} finds a median value $80$ keV for their securely inferred $\rm E_c$ measurements, fully comparable to our result.
    Since for a large fraction of our sources in our sample ($\sim50$\%)  only lower limits on the high energy cut-off are available, the values reported above are not representative of the whole sample of Swift/BAT AGN.  Therefore we use the Kaplan-Meier estimator \citep{isobe1986} in order to take the upper and lower limits into account. We find a mean value of $206\pm38$ keV.
    \citet{ricci2017} quote  a mean value of $331\pm29$ keV. The striking difference of the results possibly relates to the substantial fraction of censored values in the sample of \citet{ricci2017} (i.e. >70\%).
    
    Nevertheless, Fig. \ref{ec_dist}  suggests that the $\rm E_c$ distribution is right-skewed. Moreover the inspection of the individual errors of $\rm E_c$ parameter listed in Table \ref{spectral_fit} also reveal the presence of large and asymmetric errors for a significant fraction of the sample. Therefore the error estimation on the high energy cut-off presented  should be treated cautiously.
     Instead, we choose to estimate the probability distribution of the cut-off values. We adopt a Markov chain Monte Carlo (MCMC) method in {\sc XSPEC}, using the Goodman-Wearer algorithm, to derive the distribution of the $\rm E_c$ parameter for each accurately measured value. Then we calculate the average probability distribution of the actually measured high energy cut-offs by summing all the MCMC results of the individual sources. 

    In Fig. \ref{ec_dist} we present the average probability distribution of the $\rm E_c$ parameter for the detections after taking into account the correct error distribution for each source using the MCMC simulation approach. The corrected distribution is slightly shifted to the right due to the  uncertainties towards higher energies. The mean value of the observed $\rm E_c$ parameter, based on the MCMCs presented in Fig. \ref{ec_dist} is 150 keV. 
     The mean values of all spectral parameters are summarised in Table \ref{tab:summary}.

\section{Simulations for the estimation of the true $E_c$ distribution}

 The large number of upper limits in the spectral cut-off  may cast some doubt on the  validity of the values quoted above. The large number of upper limits can be partially attributed to the relatively limited bandwidth of the X-ray observations. Note that the value of the median (mean) cut-off is outside 
 the \nustar spectral energy band-pass. 
 In principle, the use of the survival analysis techniques alleviates this problem. However, an inherent assumption is that the  upper limits originate from the same parent population with the actually constrained values.  
 For example if there is a population with much higher coronal temperatures (cut-offs),  or if our analysis systematically identifies only  the low cut-off sources, this will not be imprinted on our results. 
 
 Moreover, a degeneracy between the photon index and the high-energy cut-off in the spectral model employed may affect the spectral fitting results \citep[e.g.][]{tortosa2018}. This degeneracy reflects the fact that an underestimated value of the $\rm E_c$ parameter can be compensated , in terms of goodness of fit, by a flatter photon index. Furthermore, imperfect modeling of the soft part of the X-ray spectrum may also induce correlations between the photon index and the high energy cut-off, particularly when  observations from different instruments are combined  \citep{molina2019} or alternatively when significant absorption is expected in the X-ray spectra \citep{balokovic2020}. 
 
 In order to determine the properties of the parent population,  we employ detailed simulations  taking into account all the possible observational degeneracies and systematics.
 
\subsection{Description of the simulation methodology}

Our idea is to simulate a number of spectral data sets with identical characteristics to those found for our sample, using however, each time, a different distribution only for the spectral cut-off $\rm E_c$ parameter. Then we repeat the same spectral analysis presented in section \ref{fitting} to derive different output distributions for the high energy cut-off parameter. 
The goal is to identify the output  $\rm E_C$ distribution,  that best matches our observed sample distribution. Then the true distribution of our population would be the known, real distribution that corresponds to the best matched simulated sample.

We  assume a skewed (Maxwell-Boltzmann) probability distribution for the
spectral cut-off parameter: 
\begin{equation}
f(x) dx =\sqrt{\frac{2}{\pi}} x^2 e^{-x^2/2\alpha}/\alpha^3 dx
\end{equation}
 
This choice is motivated from  the shape of the $\rm E_c$ distribution  in Fig.  \ref{ec_dist}. 
The distribution does not introduce unrealistic negative $\rm E_c$ values as would be the case if a Gaussian form was used.  
Moreover, this distribution presents the advantage that its shape 
is determined by only one parameter, the distribution parameter $\alpha$, thus significantly reducing the number of the 
performed simulations. The distribution parameter $\alpha$  defines its mean value through the relation $\mu= 2\alpha\sqrt{2/\pi}$.

We also assume that the true distributions of the photon index and the reflection strength of our sample follow the derived  observed distributions (see Fig. \ref{g_dist} and \ref{r_dist} respectively). This implicitly assumes that the spectral fitting does not affect the intrinsic distribution  of the $\rm \Gamma$ and R parameters. 

We simulate 14 samples assuming a different $\rm E_c$ distribution with the value of the distribution parameter $\alpha$ ranging from 25 keV to 500 keV. Each sample contains 10 times more sources than our sample. Spectra and background files are created using {\sc XSPEC} with the same count, $\rm \Gamma$ and R distributions with our sample. 
We fit the simulated data in each sample and try to measure the $\Gamma$, R and $E_c$ parameters and their uncertainties at the 90 per cent confidence level in order to compare with the observations.

 \begin{figure}
    \begin{center}
    \includegraphics[height=0.75\columnwidth]{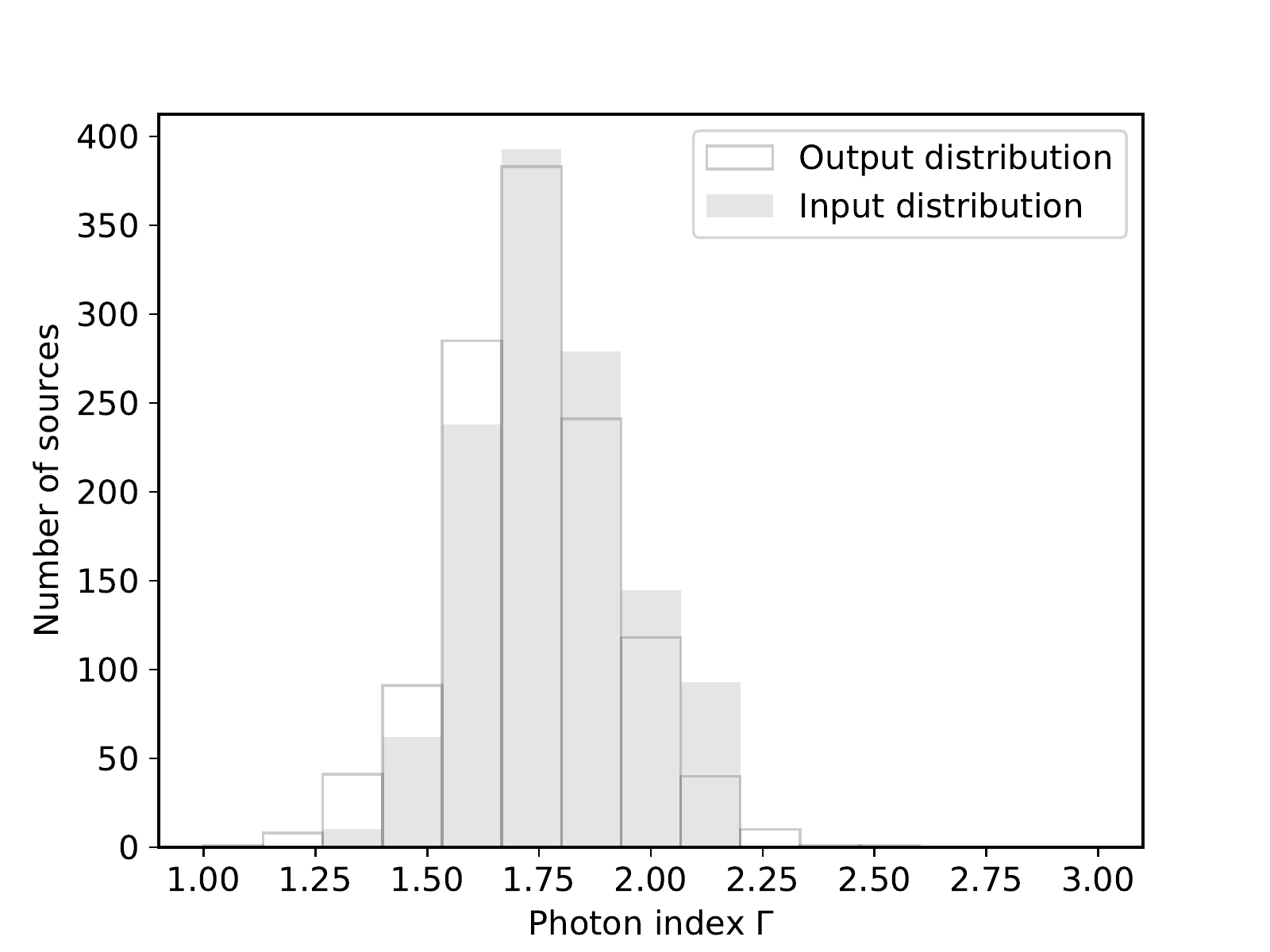}
    \end{center}
    \caption{Example of the comparison between the input and the output distributions of the photon index for a simulated sample with an $\alpha$ parameter of 175 keV.}
    \label{sim_g_dist}
    \end{figure}
    
    \begin{figure}
    \begin{center}
    \includegraphics[height=0.75\columnwidth]{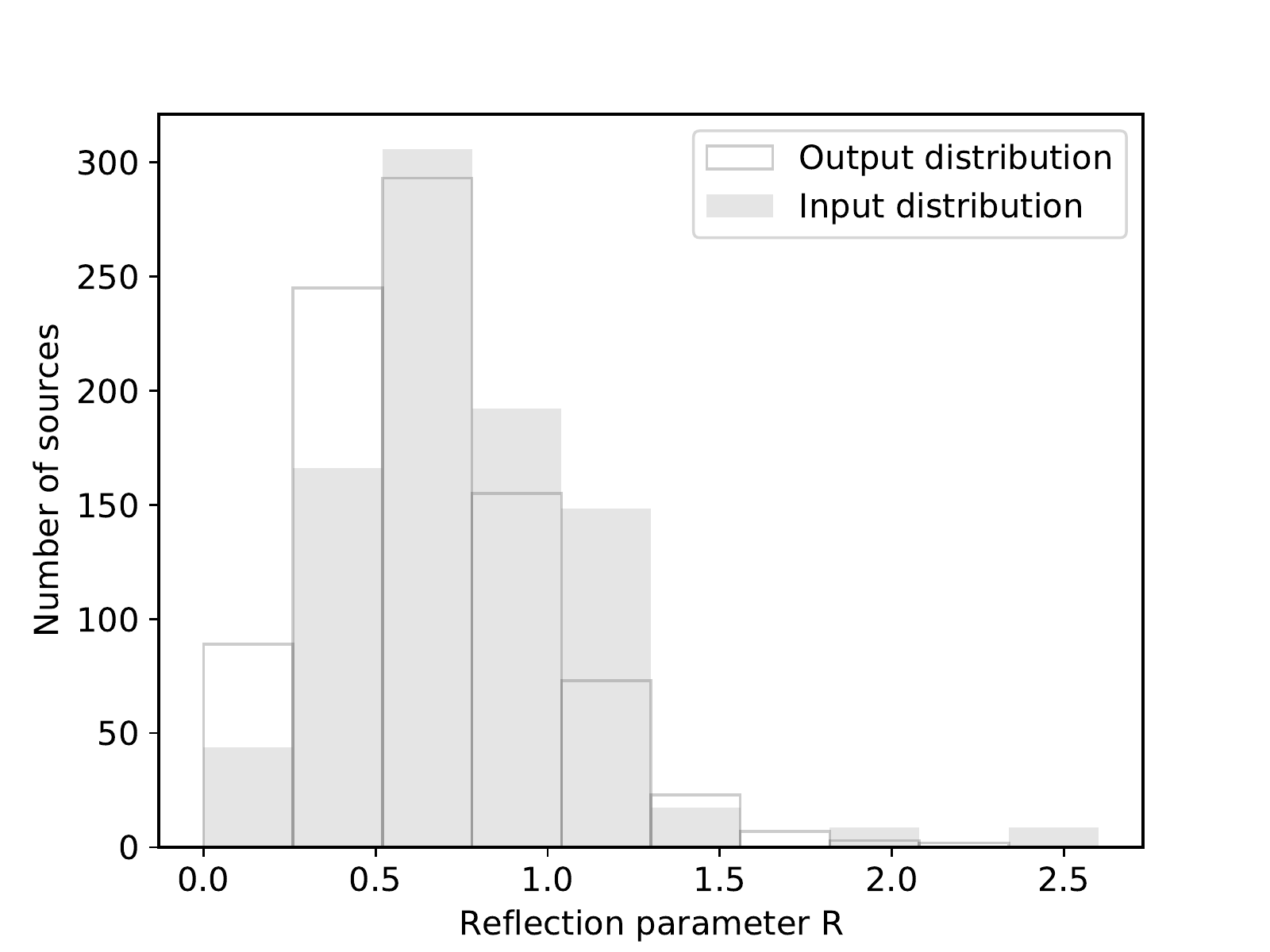}
    \end{center}
    \caption{Example of the comparison between the input and the output distributions of the reflection parameter for a simulated sample with an $\alpha$ parameter of 175 keV.}
    \label{sim_r_dist}
    \end{figure}
    
    \begin{figure*}
    \begin{center}
    \includegraphics[height=0.75\columnwidth]{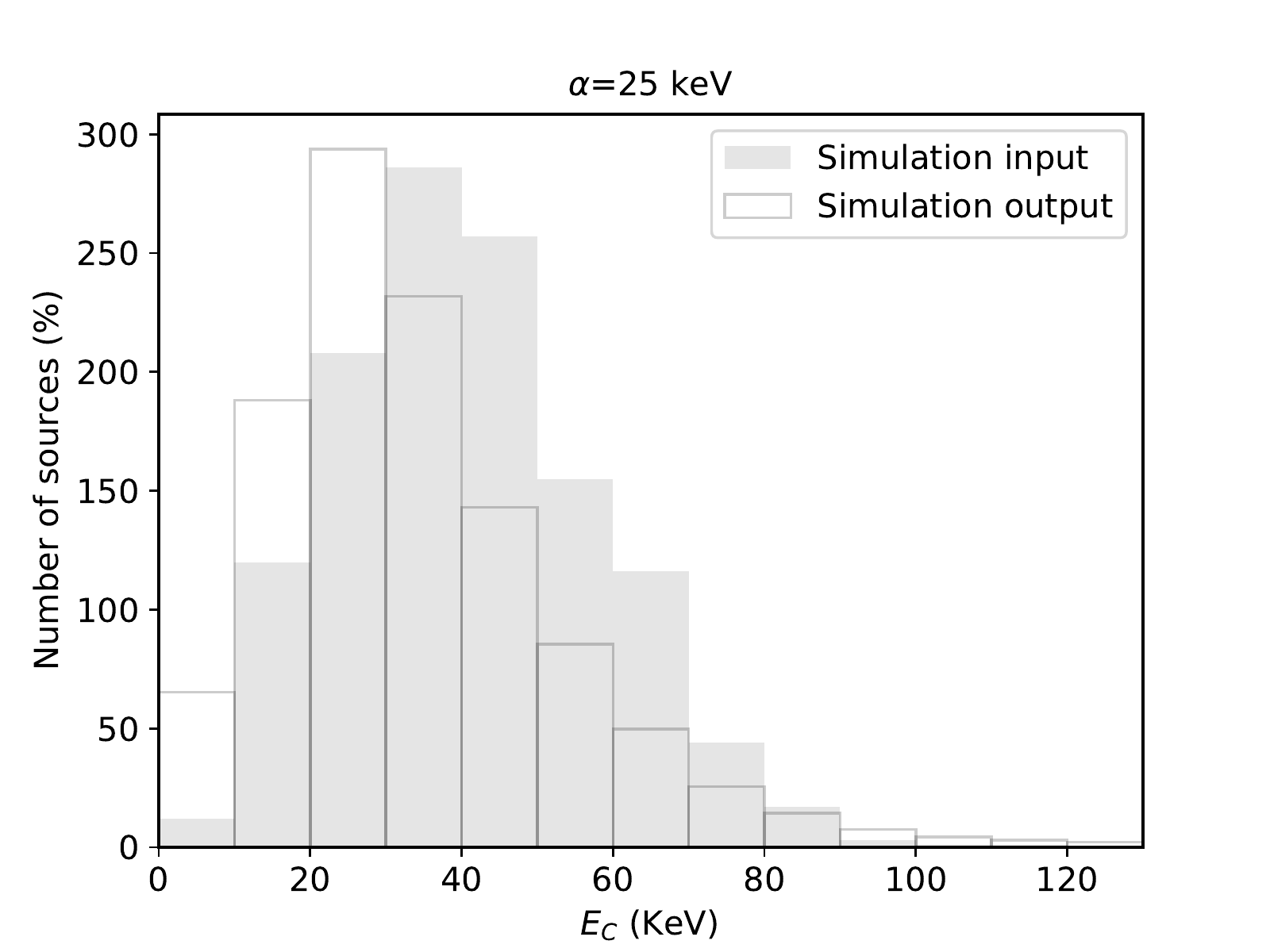}
    \includegraphics[height=0.75\columnwidth]{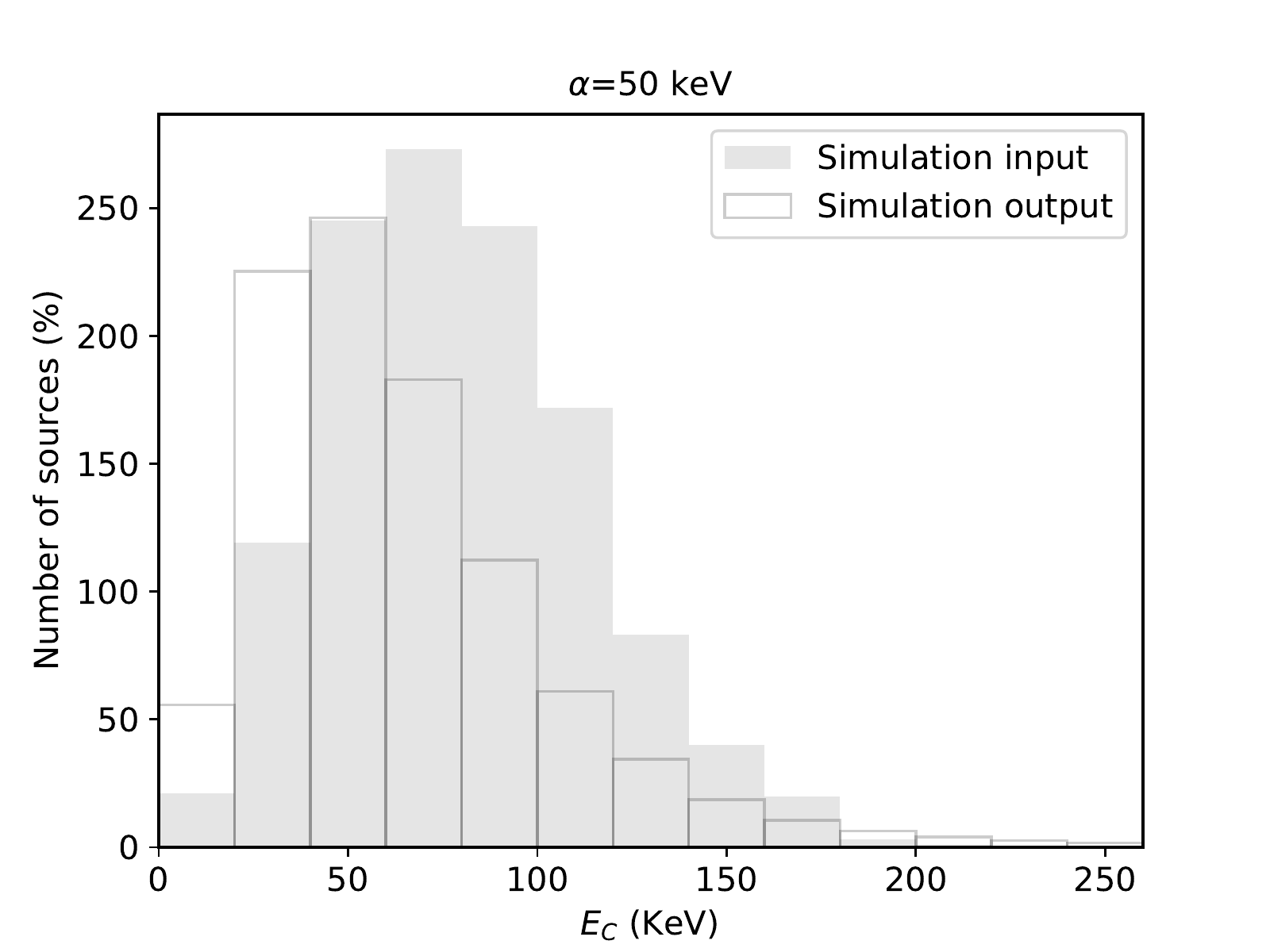}
    \includegraphics[height=0.75\columnwidth]{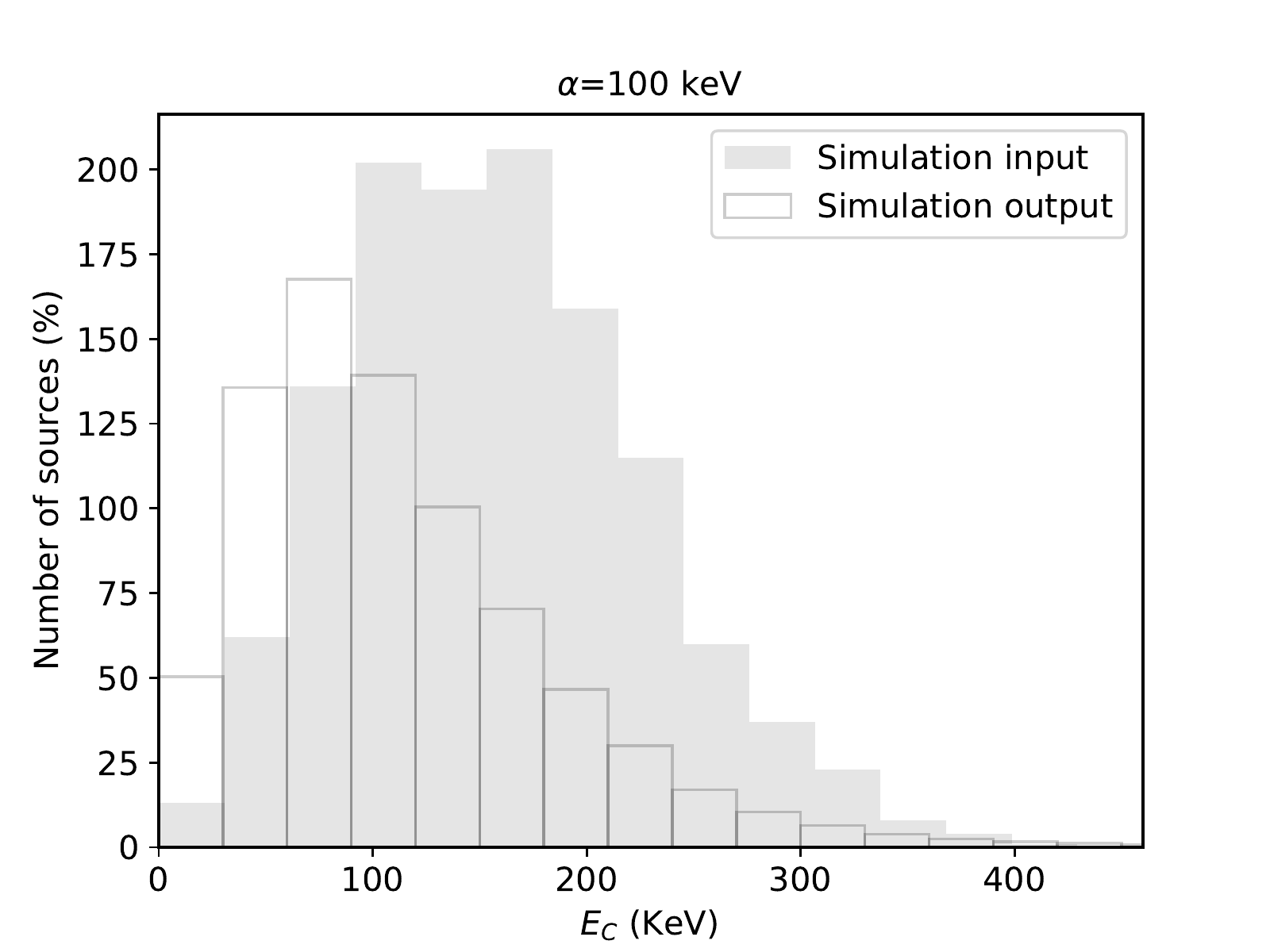}
    \includegraphics[height=0.75\columnwidth]{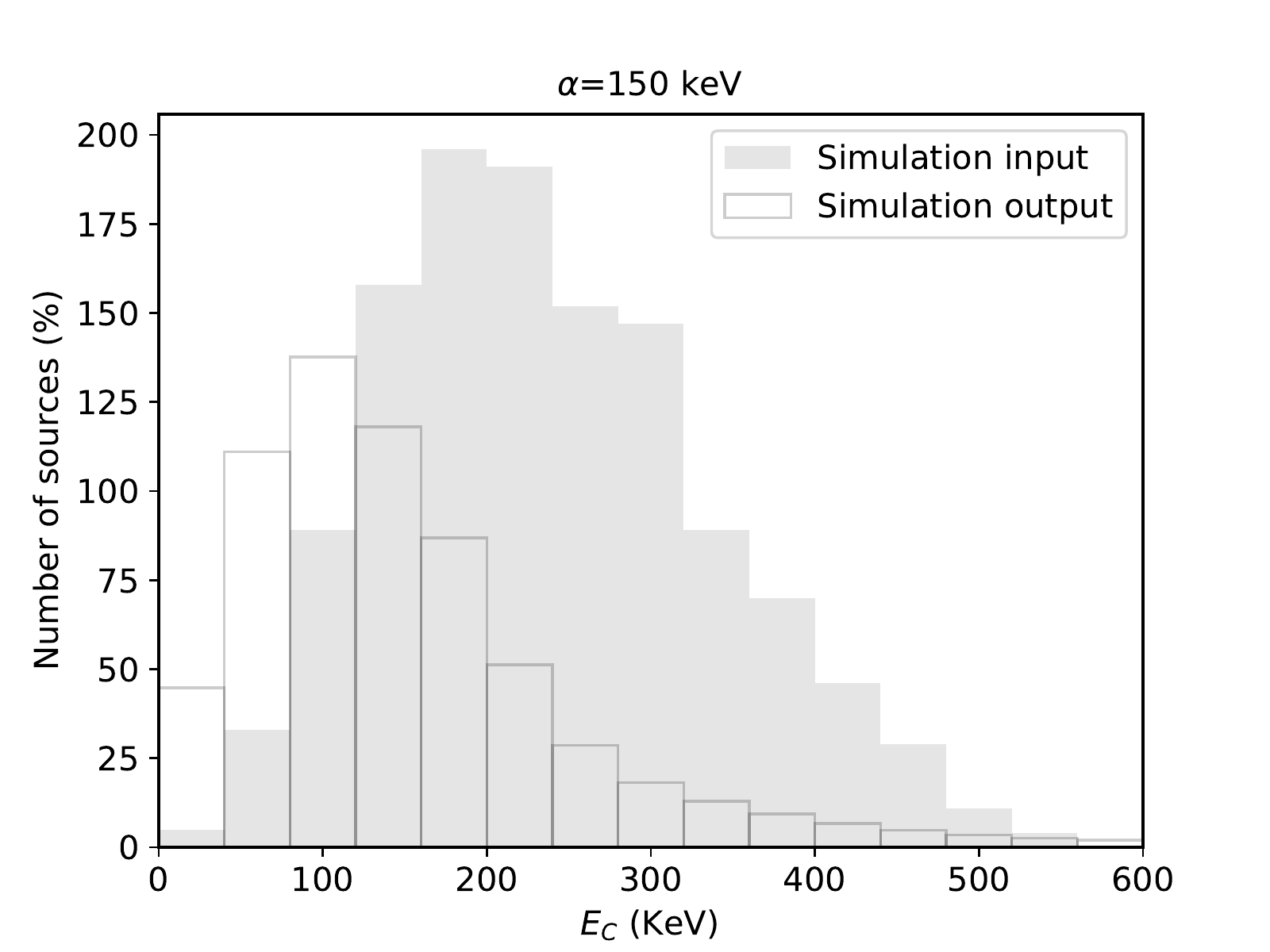}
    \includegraphics[height=0.75\columnwidth]{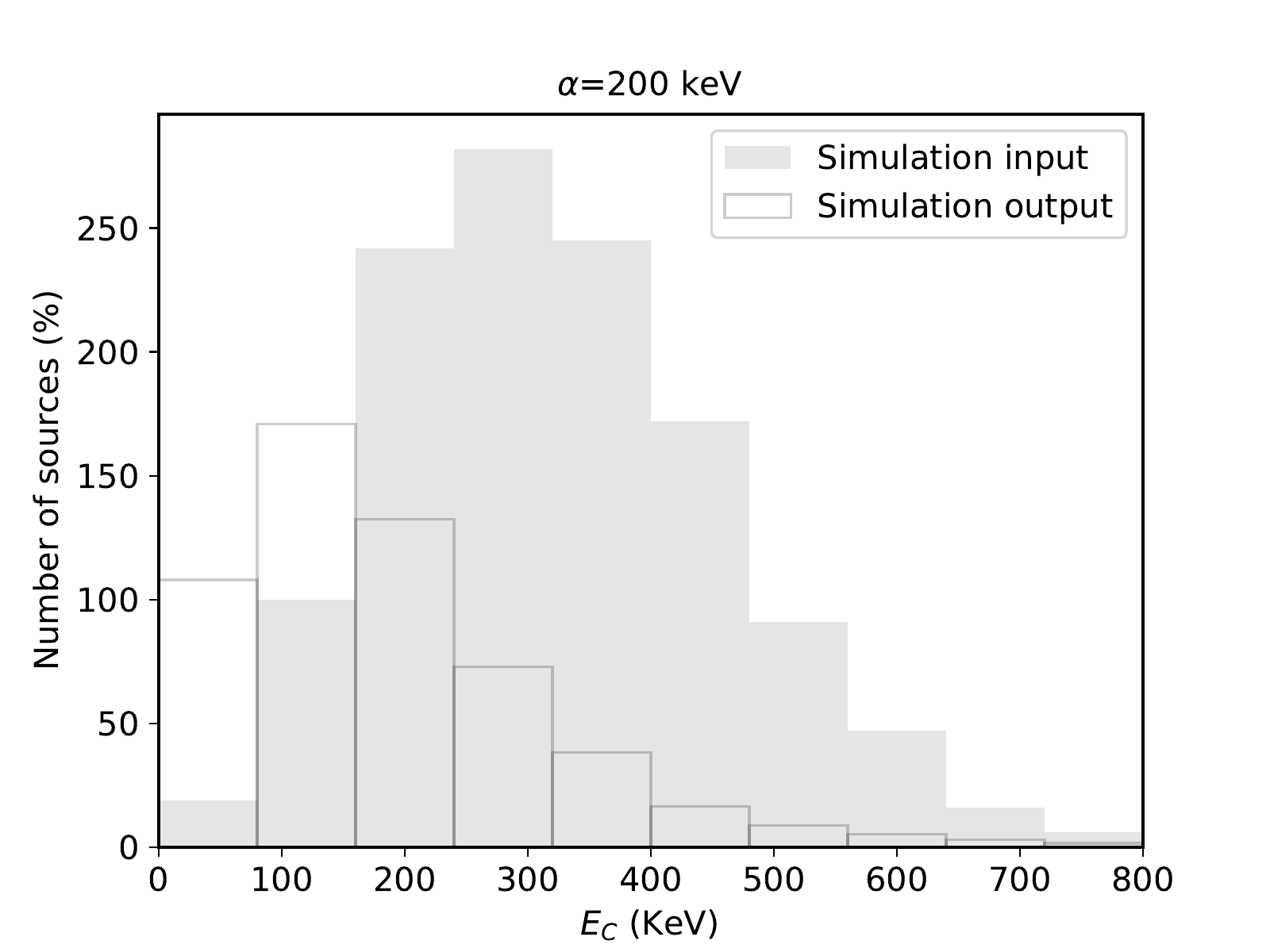}
    \includegraphics[height=0.75\columnwidth]{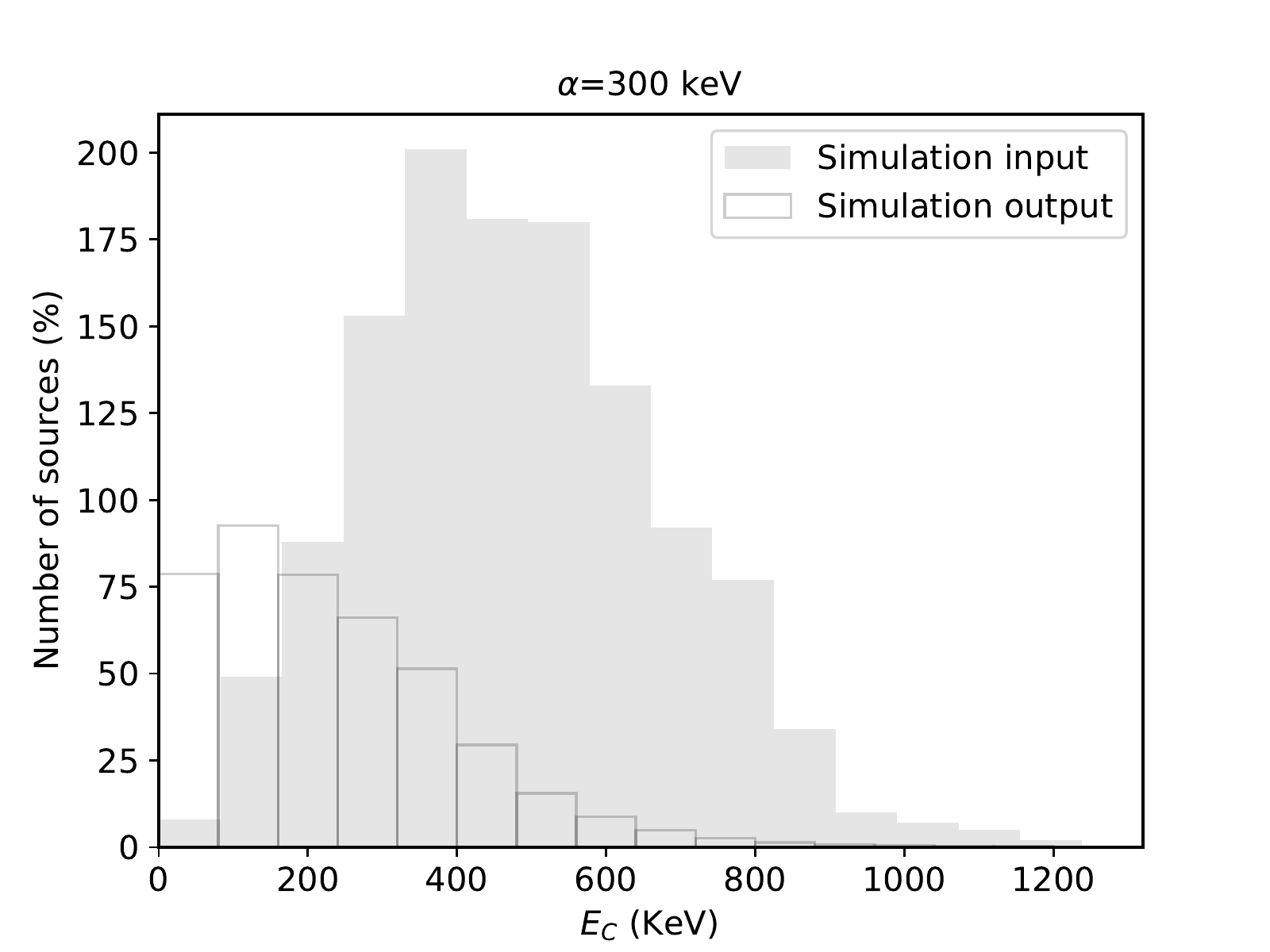}
    \end{center}
    \caption{Comparison between the input and the output distributions of the high energy cut-off parameter for a six simulated sample with an $\alpha$ parameter of 25, 50, 100, 150, 200, and 300 keV.}
    \label{sim_ec_dist}
    \end{figure*}

\subsection{Simulation Results and comparison with the data}

In all the simulations the photon index has been accurately constrained in  100 per cent of the spectral fits regardless of the assumed $\rm E_c$ distribution. The output distribution of the photon index, closely reproduces the input distribution in all cases. An example is presented in  Fig. \ref{sim_g_dist} where we plot the input and output $\Gamma$ distributions for the simulated samples with an $\alpha$ parameter of 175 keV.  The Kolmogorov-Smirnov test of the two distributions gives a probability of $\sim$0.80, supporting the hypothesis that the two distributions are equivalent. 

Similarly, the output distribution of the reflection parameter closely follows the input 
distribution. In Fig. \ref{sim_r_dist} we compare the input reflection's parameter distribution and the output one, i.e. after the spectral fitting,  for the same simulated sample presented above. Again, there is a reasonable agreement suggesting that the observed R distribution represents the true R distribution of the sample. The Kolmogorov-Smirnov statistical comparison gives a probability of $\sim$0.35, absolutely consistent with our hypothesis. We note that both $\Gamma$ and R distributions remains largely unaffected by the choice of the $E_c$ parameter. 

   \begin{figure}
    \begin{center}
    \includegraphics[height=0.75\columnwidth]{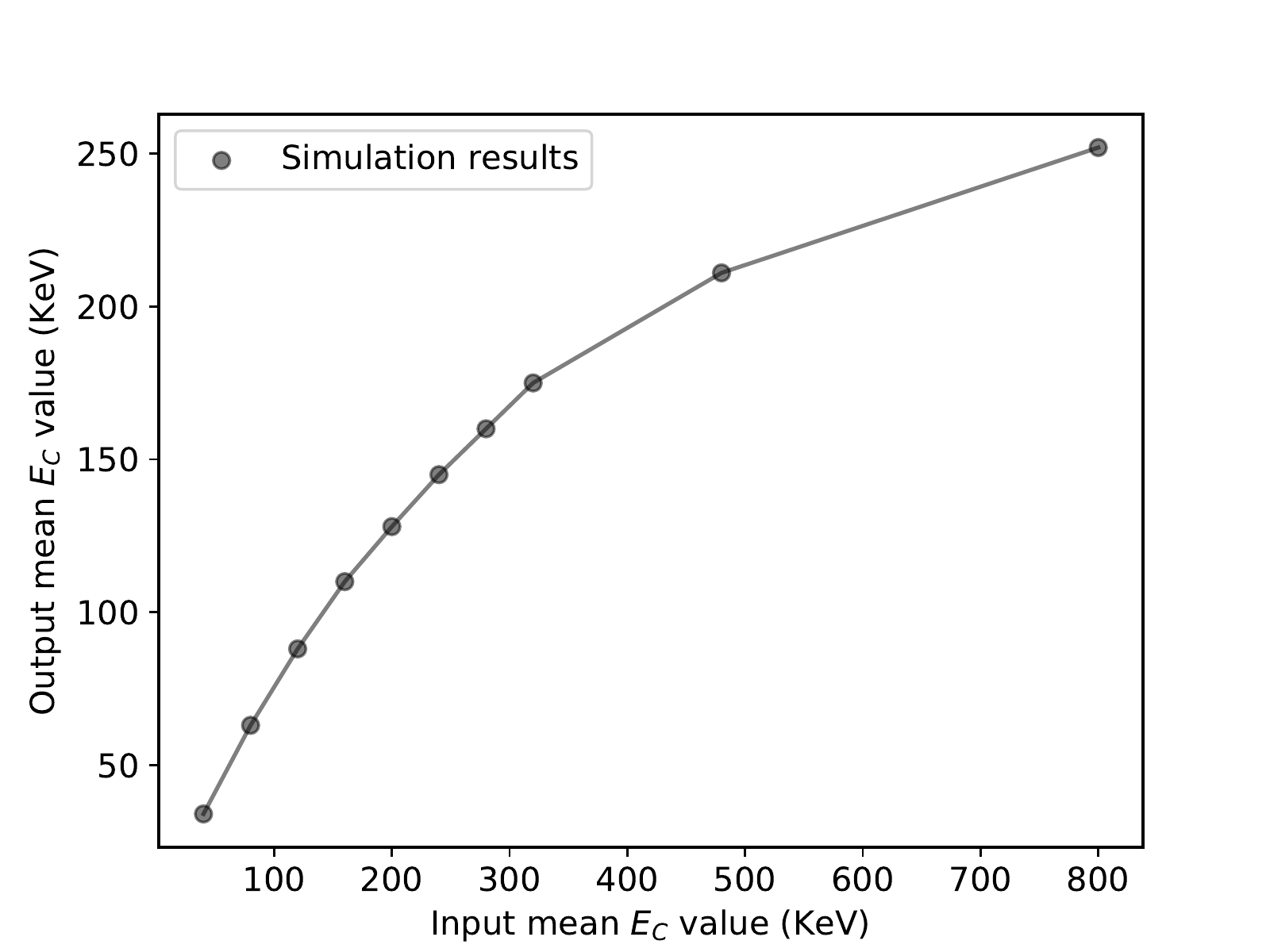}
    \end{center}
    \caption{The mean value of the spectral cut-off $\rm E_c$ of the input distribution of $\rm E_c$ versus the  output distribution.}
    \label{ec_comparison}
    \end{figure}
    
    \begin{figure}
    \begin{center}
    \includegraphics[height=0.75\columnwidth]{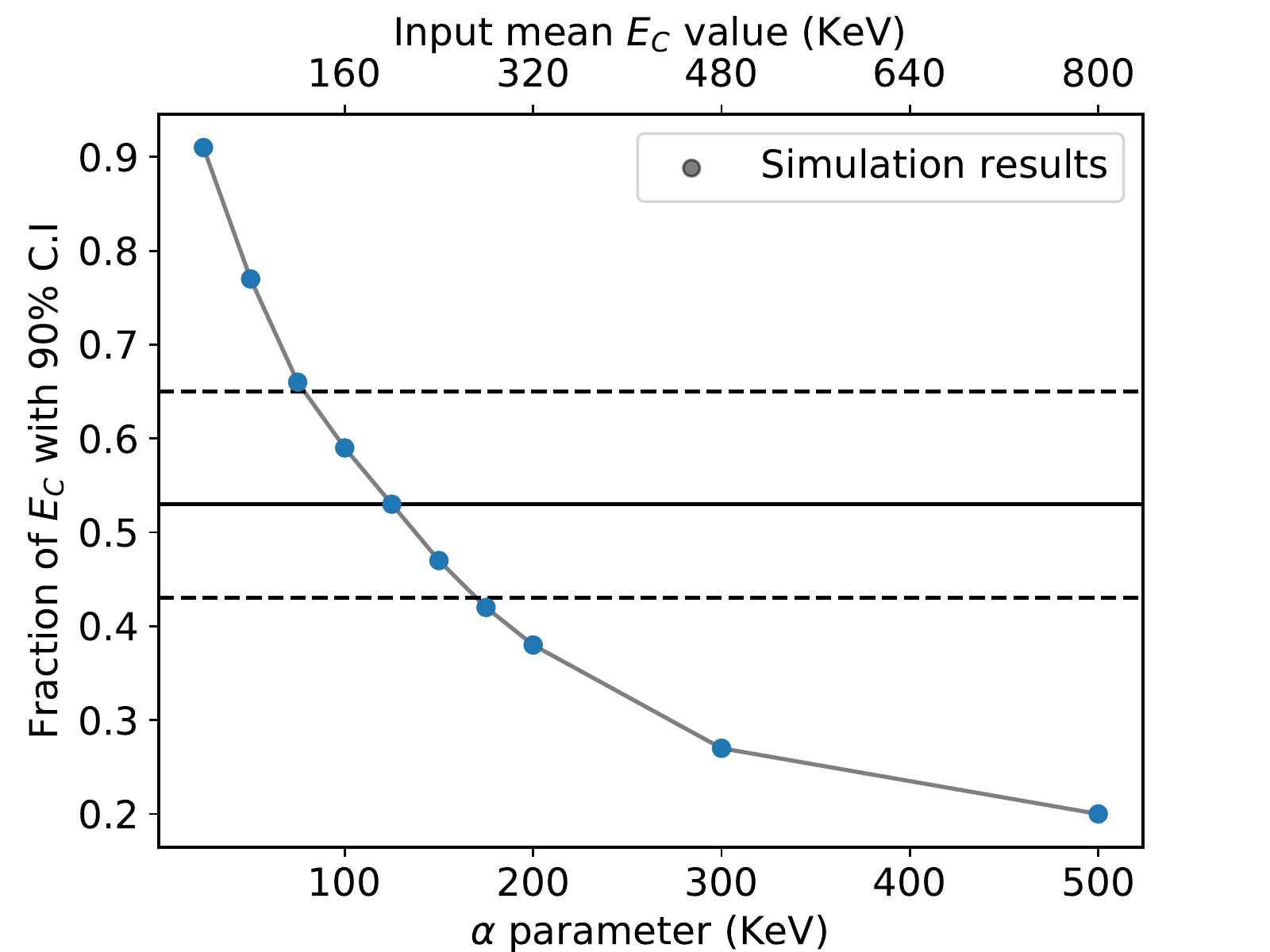}
    \end{center}
    \caption{The fraction of sources with $\rm E_c$ securely inferred values (not censored) as a function of the value of the $\rm \alpha$ parameter.
     The solid horizontal line defines the fraction of $\rm E_c$ securely inferred measurements (no censored values) while the horizontal dashed lines denote the 90 per cent uncertainty.}
    \label{det_fraction}
    \end{figure}

On the other hand the simulations reveal a quite different behaviour for the $E_c$ parameter distribution.  In Fig. \ref{sim_ec_dist} we compare the input (true) and the output (measured) distribution of the high energy cut-off, for the simulated samples, for an $\alpha$  parameter of 25, 50, 100, 150, 200 and 300 keV. Clearly, as we move towards higher $\alpha$ values the difference between the measured and the true distribution increases drastically. 
This behaviour is quantified in Fig. \ref{ec_comparison} where we plot the mean value of the input (true) distribution of $\rm E_c$ versus the mean value of the output (measured) distribution of $\rm E_c$.  We observe that when the mean $\rm E_c$ lies well within the {\it NuSTAR} pass-band, (i.e. significantly smaller than 80 keV) the output (measured) distribution is similar to the input (true) one.  This is reasonable since in this case the data can provide reliable constraints on the spectral cut-off. As we move to higher mean energies, outside the \nustar pass-band,  the input and the output mean $\rm E_c$ values diverge significantly. 

    \begin{figure*}
    \begin{center}
    \includegraphics[height=0.75\columnwidth]{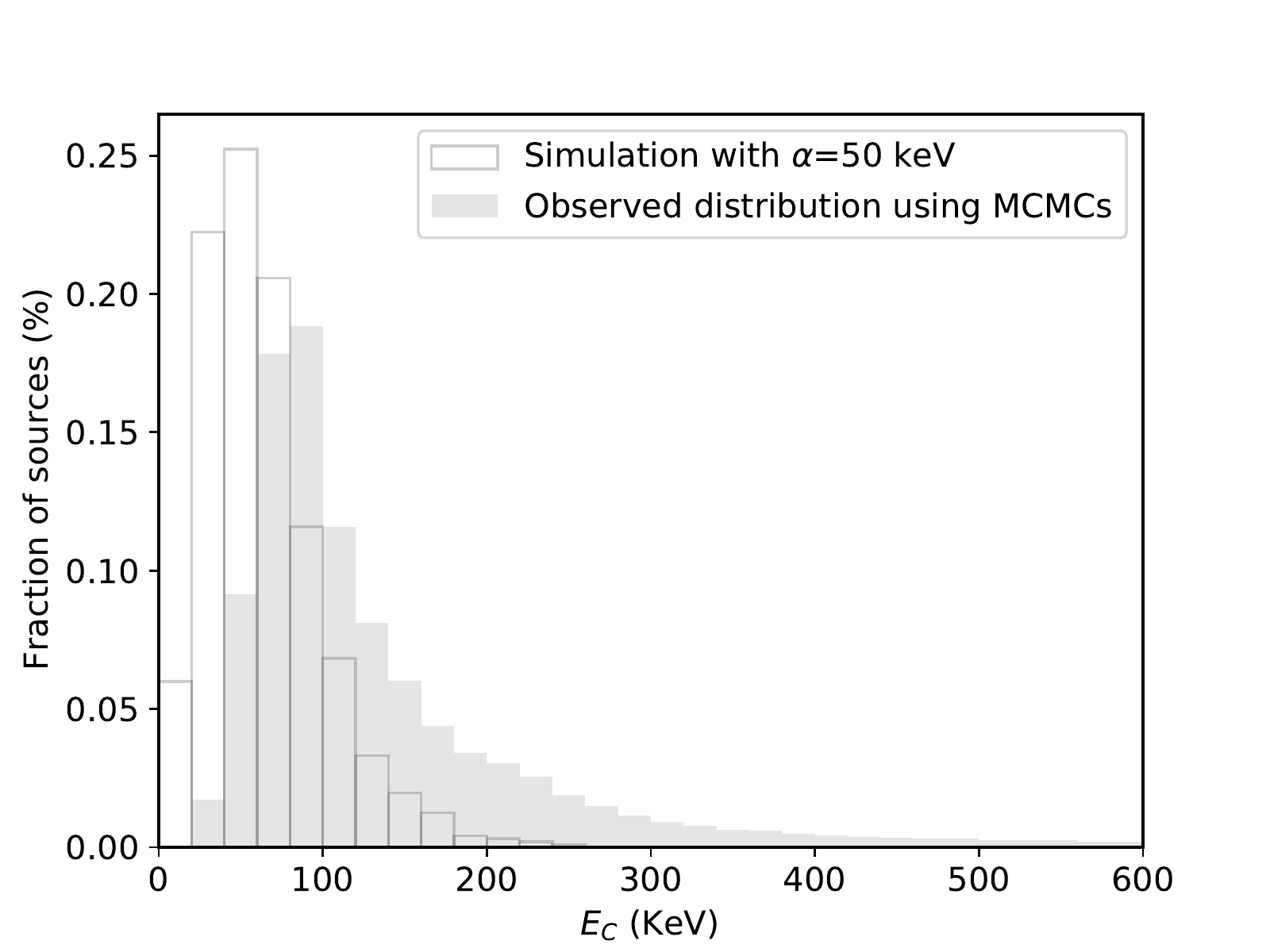}
    \includegraphics[height=0.75\columnwidth]{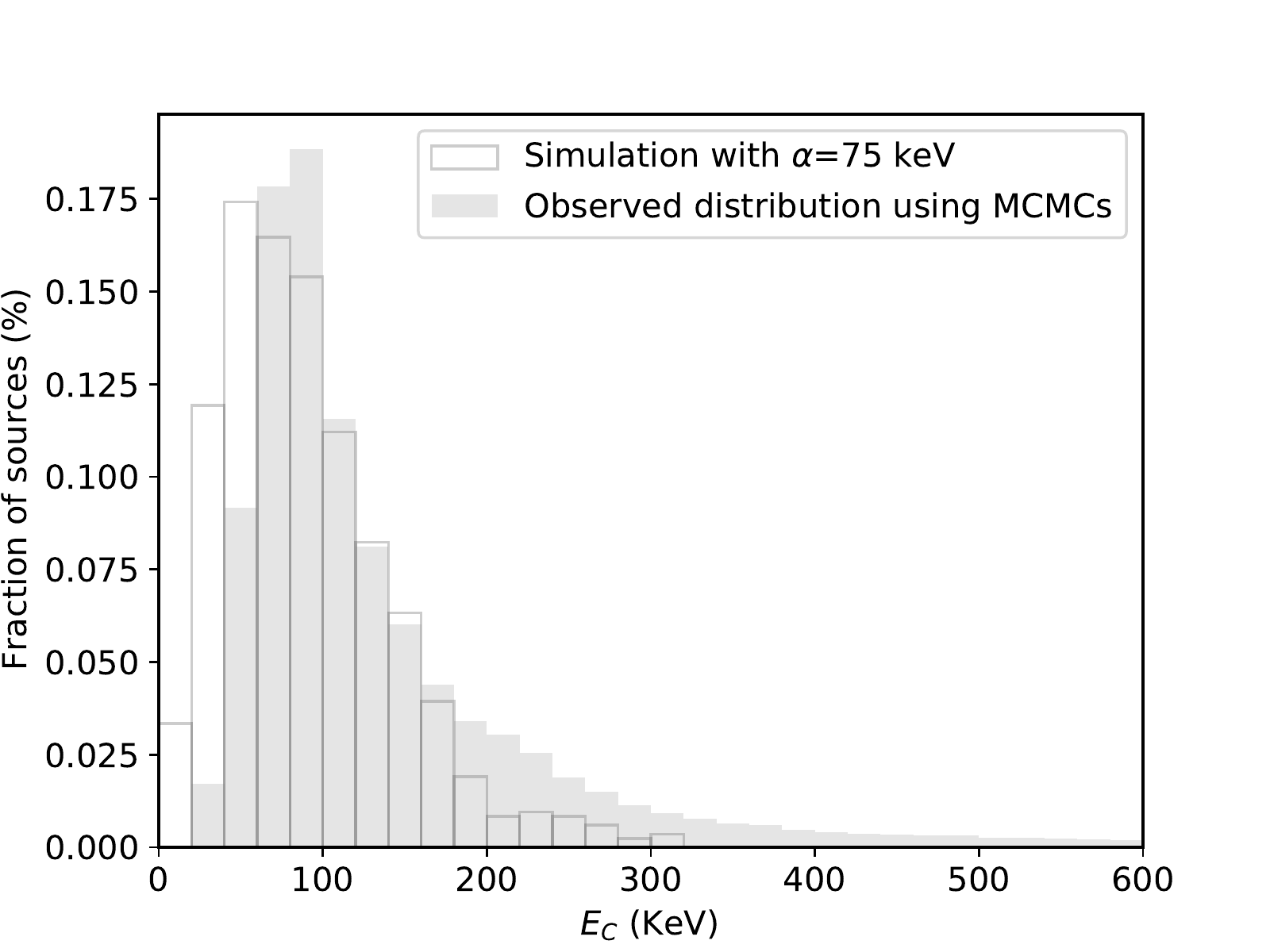}
    \includegraphics[height=0.75\columnwidth]{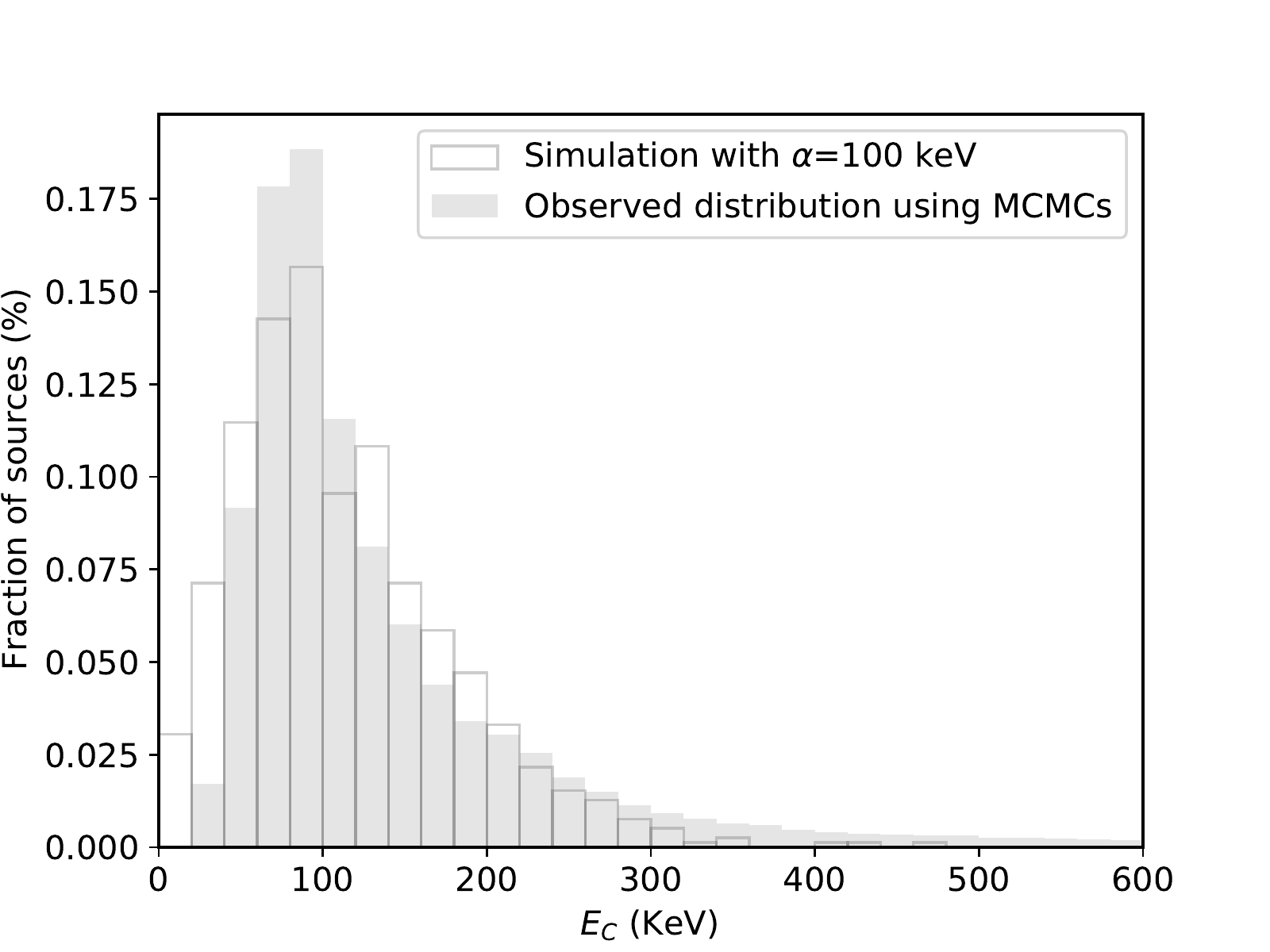}
    \includegraphics[height=0.75\columnwidth]{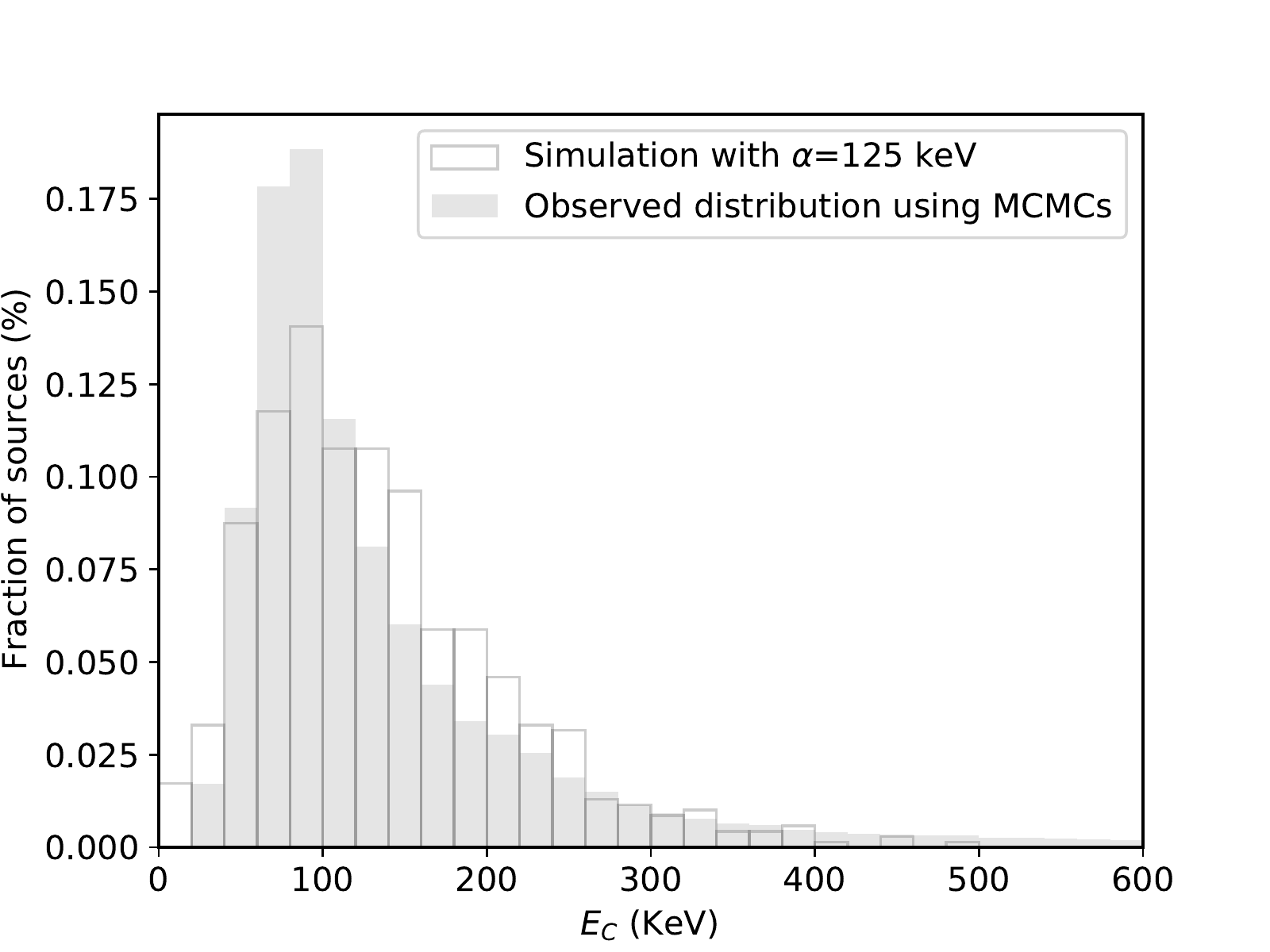}
    \includegraphics[height=0.75\columnwidth]{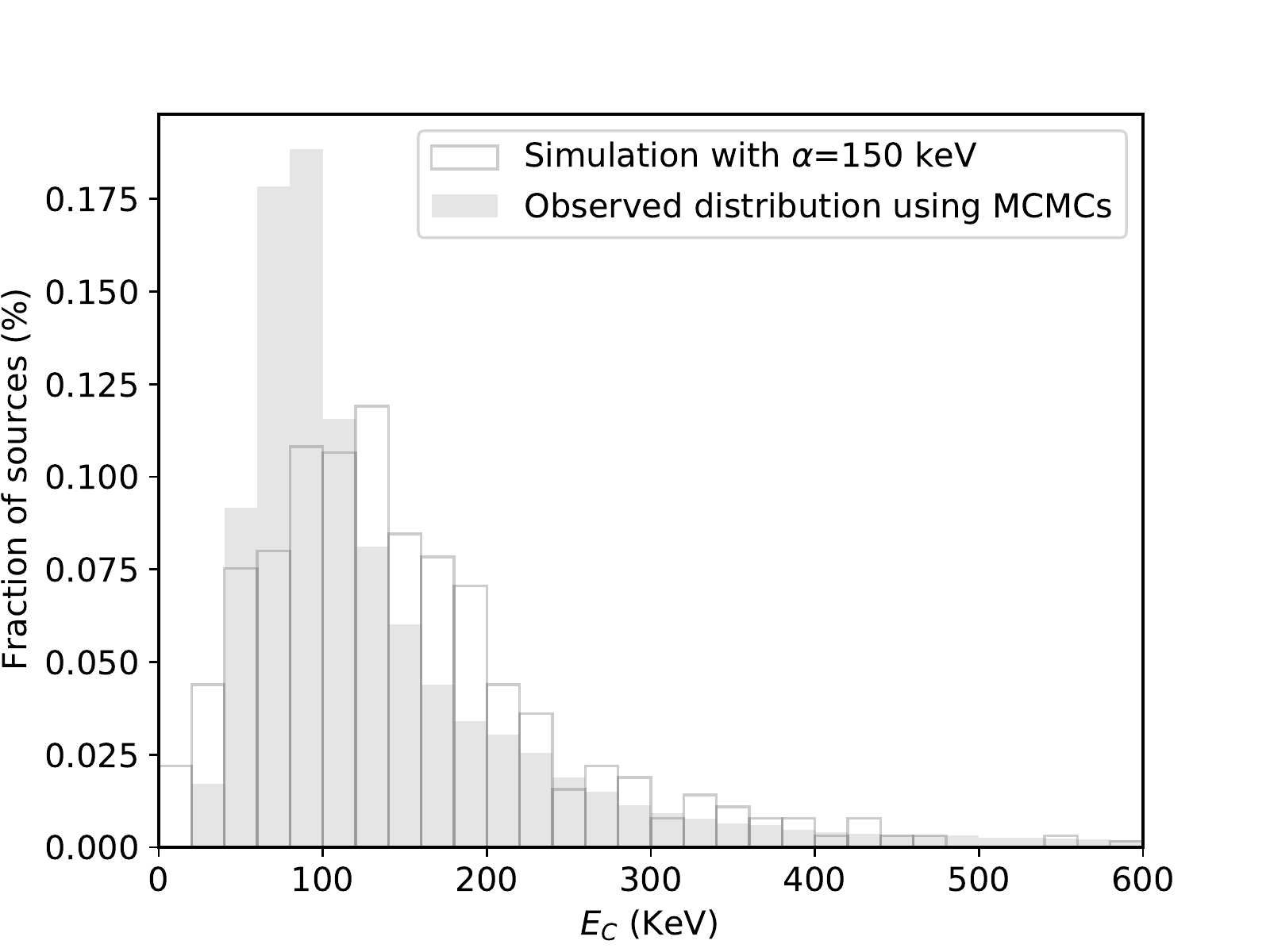}
    \includegraphics[height=0.75\columnwidth]{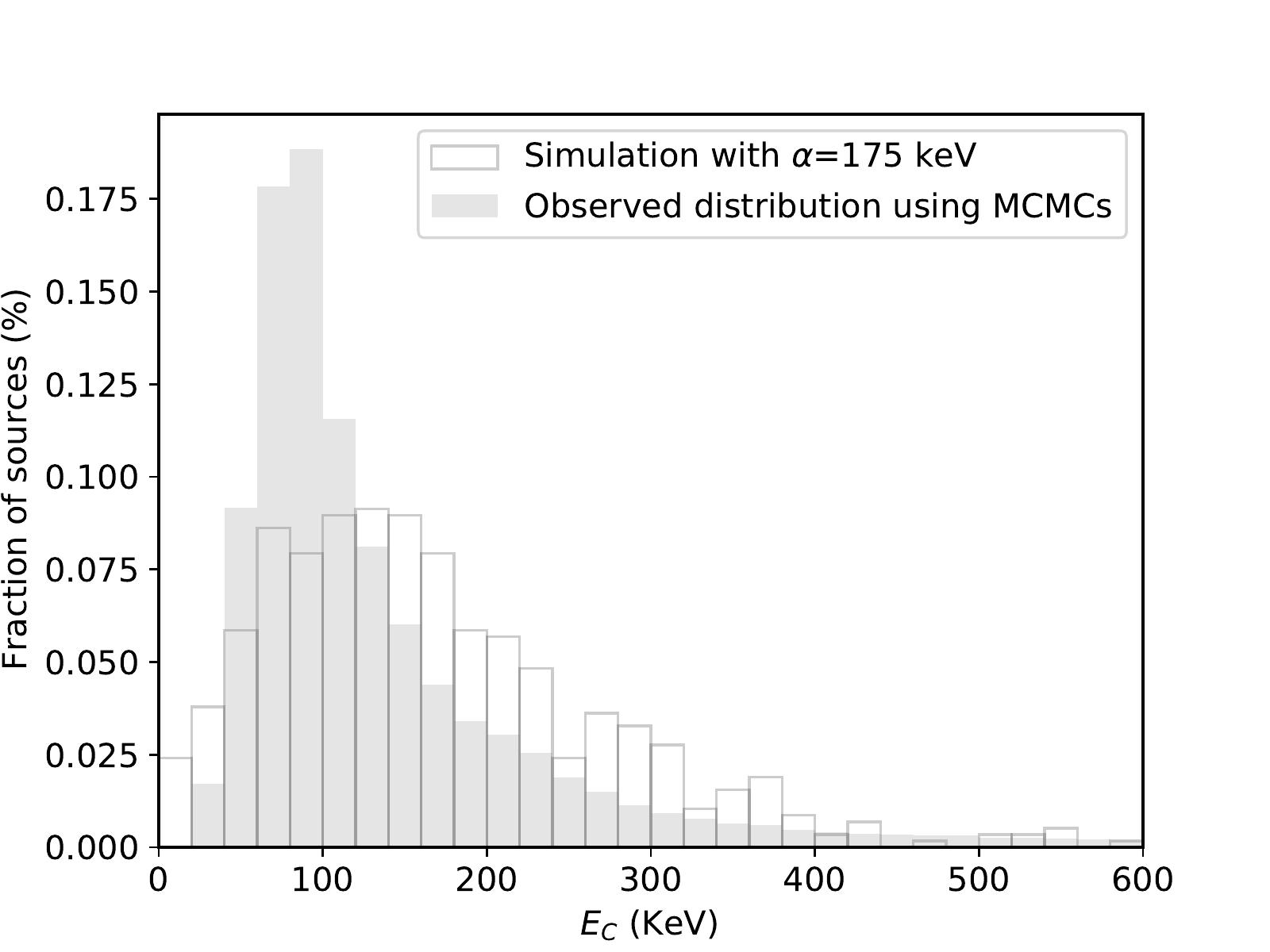}
    \end{center}
    \caption{Comparison between the observed distribution and the output distributions of the high energy cut-off parameter for a six different simulated samples with an $\alpha$ parameter of 50, 75, 100, 125, 150 and 175 keV.}
    \label{comparison_ec}
    \end{figure*}
    

In Fig. \ref{det_fraction} we plot the expected fraction of $\rm E_c$ detections (at the 90 per cent confidence interval) versus the $\rm \alpha$ parameter which characterize the input $\rm E_c$ distribution. The horizontal lines show the measured fraction of $\rm E_c$ detections in our sample (solid line) and its 90 per cent uncertainty (dashed lines). We see that the acceptable $\rm \alpha$ solution should be found within 75 keV and 165 keV with the most probable value being at 125 keV. These results correspond to a true mean value for the $\rm E_c$ parameter of $200\pm60$ keV. For values outside this range, the observed fraction of sources with well measured $\rm E_c$ values (observational data) is inconsistent with the corresponding fraction obtained from the simulations.
Furthermore, from Fig. \ref{det_fraction} we can see that the fraction of securely inferred $E_c$ values decreases with increasing $E_c$. This is also a reasonable result implying that higher $\rm E_c$ values are more difficult to be accurately measured in a given X-ray spectrum. 

More stringent constraints can be placed through the comparison of the observed distribution presented in Fig. \ref{ec_dist} with the normalized output distributions derived from the spectral fitting of the simulated samples (output distributions presented in Fig. \ref{sim_ec_dist}). In Fig. \ref{comparison_ec} we compare our  observed probability distribution with six different (output) distributions obtained from the simulations. These distributions correspond to an $\rm \alpha$ parameter of 50, 75, 100, 125, 150 and 175 keV. A Kolmogorov-Smirnov test rejects the hypothesis that the distributions are the same for all but the $\rm \alpha$=100 keV and $\rm \alpha$=125 keV solutions. This suggest a true mean value for the $E_c$ parameter of about 180 keV. This value is consistent,  albeit somewhat lower, than that derived using the survival analysis ($206\pm38$ keV).  Our simulations are based on the assumption that the parent distribution follows a Maxwell Boltzmann distribution.  However, the use of a similar symmetrical  distribution could also provide acceptable results. For example  a Gaussian distribution with mean of 200 keV and sigma=80 keV,  can also reproduce our observations. This is reasonable since this distribution is similar to a Maxwell-Boltzmann distribution with $\alpha$=125 keV.

 Combining this result with  our estimates on the mean photon index and the mean high energy cut-off, we can estimate the mean properties of the corona for the Seyfert 1 population. Following \citet[]{petrucci2001} the mean optical depth is approximately $\rm \tau_e=1.82\pm0.14$ and the mean temperature approximately $\rm kT_e=65\pm10$ keV.

\section{Summary and conclusions}

   We present the analysis of the \nustar spectra (3-80 keV) of a sample of 118 Seyfert 1 galaxies selected from the BAT all sky survey 105-month catalogue. This is the largest sample of Seyfert 1 galaxies with \nustar spectra presented in the literature. Our main goal is to constrain the cut-off energy of the power-law spectrum and hence the temperature of the hot corona that produces the X-ray emission. Our results can be summarised as follows.
   
   We find secure estimates for the spectral cut-off in 62 sources (53\% of our sample) while for the  remaining sources only lower limits could be derived. The median (mean) value for the well constrained sources is 89 (103) keV with the distribution being highly skewed towards higher energies; the 25 and 75\% quartiles are 65 and 102 keV respectively.
    As the exclusion of the lower limits bias our sample towards the detection of the low energy spectral cut-offs,  we estimate the spectral cut-off for the full sample  using survival analysis techniques. 
    The true mean value increases to $206\pm38$ keV. 
    Furthermore, we check  the validity of our results by 
    performing extensive spectral simulations.
    This is important since the derived spectral cut-off lies outside  the  \nustar spectral pass-band (3-80  keV) and this could impede the   accurate estimation of the spectral cut-off. Under the assumption of an underlying Maxwell-Boltzmann distribution, our simulations suggest  that the spectral cut-off of the parent population is $E_c\approx 160-200$ keV significantly higher than that inferred from the actual observations. This is comparable  with the value derived using survival analysis techniques. 
    
      Our work also provides strong constraints on other spectral parameters. 
      The mean value of the photon index is $\Gamma=1.78\pm0.01$. This is in good agreement with previous estimates of BAT selected Seyfert 1 galaxy spectra  which 
      have been fitted using a combination of BAT and softer X-ray spectra from {\it Chandra} {\it XMM - Newton} and \swift/XRT. 
      Also the reflection parameter derived from the full sample has a mean value of $R=0.69\pm0.04$. Again this is in reasonable agreement with the previous BAT results on Seyfert 1 galaxies but the excellent quality of the \nustar spectra reduced significantly the fraction 
      of censored values thus providing much more stringent constraints.
      
      Combining these results, our work provides stringent constraints on the mean properties of the corona. The mean optical depth is approximately $1.82\pm0.14$ and the mean temperature approximately $65\pm10$ keV.

\begin{acknowledgements}
We have made use of data from the NuSTAR mission, a project led by the California Institute of Technology, managed by the Jet Propulsion Laboratory, and funded by the National Aeronautics and Space Administration. We thank the NuSTAR Operations, Software and Calibration teams for support with the execution and analysis of these observations. This research has made use of the \nustar Data Analysis Software (NuSTARDAS) jointly developed by the Space Science Data Center (SSDC; ASI, Italy) and the California Institute of Technology (USA). This work is based on archival data, software or online services provided by the SSDC. This research has made use of the High Energy Astrophysics Science Archive Research Center Online Service, provided by the NASA/Goddard Space Flight Center and NASA’s Astrophysics Data System.
  \end{acknowledgements}


\longtab{
\begin{longtable}{rlccrl}
\caption{\label{obs_log} Log of the \nustar observations of Seyfert 1 galaxies}\\
\hline\hline
BAT obsID & Name & \nustar obsID & Redshift & Net counts & Type  \\
\hline
\endfirsthead
\caption{continued.}\\
\hline\hline
BAT obsID & Name & \nustar obsID  & Redshift &  Net counts & Type  \\
\hline
\endhead
\hline
\endfoot
 6  &   Mrk335   & 60001041005 & 0.025 &  26434   &   Sy1.2  \\
19 &   2MASXJ00341665-7905204   & 60160015002 & 0.074 &  6353   &   Sy1.0  \\
36 &   Mrk1148   & 60160028002 & 0.064 &  18300   &   Sy1.5  \\
73 &   Fairall9   & 60001130003 & 0.047 &  84991   &   Sy1.2  \\
77 &   Mrk359   & 60402021002 & 0.017 &  12822   &   Sy1.5  \\
78 &   MCG-03-04-072   & 60160061002 & 0.046 &  9359   &   Sy1.0 \\ 
98 &   1ES0152+022   & 60160080002 & 0.082 &  6202   &   Sy1.0  \\
106 &   Mrk1018   & 60301022003 & 0.042 &  2853   &   Sy1.2  \\
116 &   Mrk590   & 90201043002 & 0.026 &  7078   &   Sy1.5  \\
117 &   2MASXJ02143730-6430052   & 60061021002 & 0.074 &  5131   &   Sy1.0 \\ 
127 &   AM0224-283   & 60363002002 & 0.059 &  3060   &   Sy1.2  \\
129 &   NGC931   & 60101002004 & 0.016 &  72618   &   Sy1.5  \\
130 &   Mrk1044   & 60160109002 & 0.016 &  8144   &   Sy1.0  \\
134 &   NGC985   & 60061025002 & 0.043 &  7232   &   Sy1.5  \\
147 &   HB890241+622   & 60160125002 & 0.044 &  24951   &   Sy1.2  \\
175 &   LCRSB032315.2-420449   & 60160152002 & 0.058 &  2766   &   Sy1.0 \\ 
191 &   HE0345-3033   & 60376003002 & 0.095 &  895   &   Sy1.0  \\
214 &   3C111.0   & 60202061006 & 0.048 &  62073   &   Sy1.2  \\
216 &   NGC1566   & 80401601002 & 0.005 &  56882   &   Sy1.5  \\
220 &   1H0419-577   & 60101039002 & 0.104 &  109546   &   Sy1.5 \\ 
224 &   2MASXJ04293830-2109441   & 60260006002 & 0.070 &  1645   &   Sy1.2 \\ 
229 &   2MASXJ04372814-4711298   & 60160197002 & 0.053 &  3913   &   Sy1.0  \\
234 &   IRAS04392-2713   & 60160201002 & 0.083 &  5809   &   Sy1.5  \\
266 &   Ark120   & 60001044004 & 0.032 &  107956   &   Sy1.0 \\ 
269 &   ESO362-18   & 60201046002 & 0.012 &  51569   &   Sy1.5 \\ 
310 &   MCG+08-11-011   & 60201027002 & 0.021 &  219306   &   Sy1.5 \\ 
314 &   PKS0558-504   & 60160254002 & 0.137 &  7443   &   Sy1.0  \\
328 &   NVSSJ062335+644538   & 60376012002 & 0.086 &  4895   &   Sy1.0 \\  
376 &   1RXSJ073308.7+455511   & 60260007002 & 0.141 &  2310   &   Sy1.0 \\  
378 &   Mrk9   & 60061326002 & 0.039 &  3189   &   Sy1.2  \\
402 &   ESO209-G012   & 60160315002 & 0.040 &  11223   &   Sy1.5 \\  
409 &   PG0804+761   & 60160322002 & 0.100 &  5241   &   Sy1.0  \\
423 &   Fairall1146   & 60061082002 & 0.031 &  11993   &   Sy1.5 \\  
425 &   3C206   & 60160332002 & 0.197 &  7941   &   Sy1.0  \\
431 &   1RXSJ084521.7-353048   & 60061085002 & 0.137 &  2311   &   Sy1.0 \\  
447 &   IRAS09149-6206   & 90401630002 & 0.057 &  67745   &   Sy1.0  \\
449 &   Mrk704   & 60061090002 & 0.029 &  9936   &   Sy1.2  \\
455 &   MCG+04-22-042   & 60061092002 & 0.032 &  17273   &   Sy1.2 \\  
458 &   Mrk110   & 60201025002 & 0.035 &  292836   &   Sy1.5  \\
473 &   3C227   & 60061329002 & 0.085 &  8066   &   Sy1.2  \\
485 &   ESO374-G025   & 60160384002 & 0.023 &  1556   &   Sy1.0 \\  
495 &   2MASXJ10195855-0234363   & 60260015002 & 0.060 &  550   &   Sy1.0 \\  
497 &   NGC3227   & 60202002002 & 0.003 &  83393   &   Sy1.5  \\
512 &   SDSSJ104326.47+110524.2   & 60376004002 & 0.047 &  6699   &   Sy1.5 \\  
524 &   Mrk728   & 60061338002 & 0.035 &  4050   &   Sy1.5  \\
530 &   NGC3516   & 60002042004 & 0.008 &  20304   &   Sy1.2  \\
532 &   IC2637   & 60061208002 & 0.029 &  9471   &   Sy1.5  \\
542 &   ARP151   & 60160430002 & 0.021 &  2410   &   Sy1.2  \\
552 &   Mrk739E   & 60260008002 & 0.029 &  3416   &   Sy1.0  \\
556 &   SBS1136+594   & 60160443002 & 0.060 &  10380   &   Sy1.5 \\ 
558 &   NGC3783   & 60101110004 & 0.009 &  70805   &   Sy1.2  \\
565 &   KUG1141+371   & 60160449002 & 0.038 &  7438   &   Sy1.2  \\ 
566 &   UGC06728   & 60376007002 & 0.006 &  34838   &   Sy1.2  \\
567 &   2MASXJ11454045-1827149   & 60302002006 & 0.032 &  26050   &   Sy1.2 \\ 
574 &   2MASXJ11491868-0416512   & 60061215002 & 0.084 &  5928   &   Sy1.0  \\
576 &   PG1149-110   & 60160458002 & 0.049 &  1227   &   Sy1.2  \\
583 &   Mrk1310   & 60160465002 & 0.019 &  7840   &   Sy1.5  \\
585 &   NGC4051   & 60401009002 & 0.002 &  217209   &   Sy1.5  \\
587 &   GQCom   & 60160469002 & 0.165 &  4610   &   Sy1.5  \\
589 &   2MASXJ12055599+4959561   & 60061357002 & 0.063 &  1666   &   Sy1.5  \\ 
608 &   Mrk766   & 60001048002 & 0.013 &  74065   &   Sy1.5  \\
611 &   Mrk205   & 60160490002 & 0.071 &  7215   &   Sy1.0  \\
623 &   Mrk771   & 60061229002 & 0.063 &  3368   &   Sy1.2  \\
631 &   NGC4593   & 60001149002 & 0.009 &  23920   &   Sy1.0  \\
636 &   WKK1263   & 60160510002 & 0.024 &  21298   &   Sy1.5  \\
644 &   6dFJ1254564-265702   & 60363001002 & 0.059 &  4877   &   Sy1.2 \\ 
686 &   NGC5273   & 60061350002 & 0.003 &  16637   &   Sy1.5  \\
690 &   2MASSiJ1346085+732053   & 60160556002 & 0.290 &  2194   &   Sy1.0 \\ 
694 &   IC4329A   & 60001045002 & 0.016 &  697782   &   Sy1.5  \\
695 &   UM614   & 60160560002 & 0.032 &  5760   &   Sy1.5  \\
697 &   Mrk279   & 60160562002 & 0.03 &  27192   &   Sy1.5  \\
717 &   NGC5548   & 60002044006 & 0.017 &  85416   &   Sy1.5  \\
726 &   Mrk813   & 60160583002 & 0.110 &  8088   &   Sy1.2  \\
728 &   Mrk1383   & 60061254002 & 0.086 &  10981   &   Sy1.0 \\ 
730 &   Mrk684   & 60160586002 & 0.046 &  2867   &   Sy1.0  \\
735 &   Mrk817   & 60160590002 & 0.031 &  7459   &   Sy1.2  \\
741 &   IGRJ14471-6414   & 60061257002 & 0.053 &  2608   &   Sy1.0  \\
750 &   WKK4438   & 60401022002 & 0.016 &  35926   &   Sy1.5  \\
753 &   Mrk841   & 60101023002 & 0.036 &  16341   &   Sy1.2  \\
754 &   Mrk1392   & 60160605002 & 0.036 &  4818   &   Sy1.5  \\
765 &   2MASXJ15144217-8123377   & 60061263002 & 0.068 &  2820   &   Sy1.2 \\ 
774 &   Mrk290   & 60061266004 & 0.029 &  8903   &   Sy1.5  \\
794 &   LEDA100168   & 60160631002 & 0.183 &  3360   &   Sy1.0 \\ 
795 &   WKK6092   & 60160632002 & 0.015 &  9800   &   Sy1.5  \\
810 &   VIIZw653   & 60160639002 & 0.063 &  6125   &   Sy1.2  \\
815 &   Mrk885   & 60160641002 & 0.025 &  3725   &   Sy1.0  \\
833 &   2MASXJ16481523-3035037   & 60160648002 & 0.031 &  1956   &   Sy1.0 \\ 
888 &   2MASXJ17311341+1442561   & 60161666002 & 0.082 &  2124   &   Sy1.0  \\
905 &   2E1739.1-1210   & 60160670002 & 0.037 &  10453   &   Sy1.0  \\
907 &   4C+18.51   & 60160672002 & 0.186 &  2980   &   Sy1.0  \\
912 &   1RXSJ174538.1+290823   & 60160674002 & 0.110 &  6256   &   Sy1.5 \\ 
924 &   Mrk507   & 60160675002 & 0.056 &  810   &   Sy1.2  \\
925 &   VIIZw742   & 60160676002 & 0.063 &  1881   &   Sy1.2 \\ 
948 &   2MASXiJ1802473-145454   & 60160680002 & 0.036 &  17794   &   Sy1.2 \\ 
967 &   H1821+643   & 60160683002 & 0.297 &  13180   &   Sy1.2  \\
976 &   1RXSJ182557.5-071021   & 60160688002 & 0.037 &  1076   &   Sy1.0 \\ 
984 &   3C382   & 60001084002 & 0.058 &  102343   &   Sy1.2  \\
994 &   3C390.3   & 60001082003 & 0.056 &  81788   &   Sy1.5  \\
1013 &   2MASXJ18560128+1538059   & 60160701002 & 0.084 &  7696   &   Sy1.2 \\ 
1021 &   CGCG229-015   & 60160705002 & 0.028 &  4594   &   Sy1.0  \\
1032 &   ESO141-G055   & 60201042002 & 0.036 &  97395   &   Sy1.2  \\
1041 &   2MASSJ19334715+3254259   & 60160714002 & 0.058 &  7752   &   Sy1.2 \\ 
1042 &   2MASXJ19373299-0613046   & 60101003002 & 0.010 &  54586   &   Sy1.5  \\
1043 &   2MASXJ19380437-5109497   & 60160716002 & 0.040 &  7531   &   Sy1.2  \\
1046 &   NGC6814   & 60201028002 & 0.005 &  215189   &   Sy1.5  \\
1082 &   4C+21.55   & 60160740002 & 0.173 &  7435   &   Sy1.0  \\
1084 &   2MASXJ20350566+2603301   & 60160741002 & 0.048 &  2017   &   Sy1.5  \\
1088 &   4C+74.26   & 60001080006 & 0.104 &  106864   &   Sy1.0  \\
1090 &   Mrk509   & 60101043002 & 0.034 &  323647   &   Sy1.2  \\
1106 &   2MASXJ21192912+3332566   & 60061358002 & 0.051 &  8248   &   Sy1.5 \\ 
1111 &   SWIFTJ212745.6+565636   & 60001110005 & 0.014 &  95594   &   Sy1.0  \\
1117 &   2MASXJ21355399+4728217   & 60160761002 & 0.025 &  7267   &   Sy1.5  \\
1125 &   RXJ2145.5+1102   & 60160767002 & 0.209 &  2093   &   Sy1.0  \\
1162 &   MCG+01.57-016   & 60061343002 & 0.025 &  4026   &   Sy1.5  \\
1172 &   MR2251-178   & 60102025004 & 0.064 &  52719   &   Sy1.2  \\
1182 &   NGC7469   & 60101001008 & 0.016 &  28817   &   Sy1.5  \\
1381 &   PG1100+772   & 60463031002 & 0.311 &  3196   &   Sy1.0  \\
1471 &   Mrk871   & 60361014002 & 0.033 &  4931   &   Sy1.0  \\
1503 &   Mrk506   & 60464138002 & 0.043 &  3421   &   Sy1.5  \\
1536 &   2MASSIJ1830231+731310   & 60464150002 & 0.123 &  6803   &   Sy1.0 \\ 
1581 &   3C410   & 60463069002 & 0.248 &  2236   &   Sy1.5  \\
1611 &   ESO344-G016   & 60361017002 & 0.039 &  9766      &   Sy1.5 \\ 

\hline
\end{longtable}
}

\longtab{
\begin{longtable}{cccccccccc}
\caption{\label{spectral_fit} Log of the \nustar observations of Seyfert 1 galaxies}\\
\hline\hline
BAT obsID &  $\rm \Gamma^{~(a)}$ & $\rm R^{(a)}$ & $\rm E_c^{(b)}$ & $\rm F_{2-10~keV}^{~(c)}$ & $\rm L_{2-10~keV}^{(d)}$ & $\rm F_{20-40~keV}^{~(c)}$ & $\rm L_{20-40~keV}^{(d)}$ & $\rm \chi^2$/dof & error $\rm id^{~(e)}$  \\
\hline
\endfirsthead
\caption{continued.}\\
\hline\hline
BAT obsID &  $\rm \Gamma^a$ & $\rm R^a$ & $\rm E_c^b$(keV) & $\rm F_{2-10~keV}^{~a}$ & $\rm L_{2-10~keV}^{b}$ & $\rm F_{20-40~keV}^{~a}$ & $\rm L_{20-40~keV}^{b}$ & $\rm \chi^2$/dof & error $\rm id^{~(e)}$  \\
\hline
\endhead
\hline
\endfoot
6 & 2.02 $^{+0.05 }_{ -0.06 }$ & 1.90 $^{+ 0.34 }_{ -0.31 }$ & >365  & 0.65 & 0.1 & 0.53 & 0.08 & 795 / 715 & 0 \\
19 & 2.03 $^{+0.14 }_{ -0.14 }$ & 1.29 $^{+ 0.64 }_{ -0.56 }$ & 60 $^{+ 77 }_{ -23 }$ & 0.78 & 1.05 & 0.31 & 0.43 & 259 / 259 & 1 \\
36 & 1.78 $^{+0.07 }_{ -0.07 }$ & 0.14 $^{+ 0.18 }_{ -0.13 }$ & 87 $^{+ 76 }_{ -28 }$ & 2.05 & 2.01 & 1.04 & 1.04 & 575 / 547 & 1 \\
73 & 1.94 $^{+0.03 }_{ -0.03 }$ & 0.85 $^{+ 0.12 }_{ -0.12 }$ & 196 $^{+ 112 }_{ -54 }$ & 2.45 & 1.28 & 1.38 & 0.72 & 1014 / 1051 & 1 \\
77$^{\star}$ & 1.87 $^{+0.08 }_{ -0.08 }$ & 0.89 $^{+ 0.32 }_{ -0.28 }$ & >118 & 0.61 & 0.04 & 0.41 & 0.03 & 492 / 481 & 0 \\
78 & 1.82 $^{+0.09 }_{ -0.09 }$ & 0.68 $^{+ 0.34 }_{ -0.29 }$ & 102 $^{+ 169 }_{ -42 }$ & 1.08 & 0.53 & 0.66 & 0.33 & 358 / 362 & 1 \\
98$^{\star}$ & 1.75 $^{+0.11 }_{ -0.12 }$ & 0.34 $^{+ 0.38 }_{ -0.21 }$ & >84 & 0.68 & 1.11 & 0.46 & 0.77 & 277 / 259 & 0 \\
106$^{\star}$ & 1.85 $^{+0.22 }_{ -0.21 }$ & 0.85 $^{+ 0.76 }_{ -0.58 }$ & >47 & 0.17 & 0.07 & 0.1 & 0.04 & 152 / 152 & 0 \\
116 & 1.71 $^{+0.13 }_{ -0.13 }$ & 1.16 $^{+ 0.5 }_{ -0.42 }$ & 66 $^{+ 86 }_{ -26 }$ & 0.32 & 0.05 & 0.24 & 0.04 & 313 / 307 & 1 \\
117 & 1.69 $^{+0.13 }_{ -0.09 }$ & <0.34 & 99 $^{+ 235 }_{ -53 }$ & 0.54 & 0.72 & 0.32 & 0.42 & 211 / 219 & 1 \\
127 & 1.81 $^{+0.14 }_{ -0.17 }$ & 0.83 $^{+ 0.72 }_{ -0.54 }$ & >80 & 0.31 & 0.26 & 0.27 & 0.23 & 127 / 148 & 0 \\
129 & 1.83 $^{+0.03 }_{ -0.03 }$ & 0.71 $^{+ 0.12 }_{ -0.11 }$ & 172 $^{+ 108 }_{ -50 }$ & 2.92 & 0.18 & 1.89 & 0.12 & 1027 / 1061 & 1 \\
130 & 2.12 $^{+0.09 }_{ -0.11 }$ & 1.13 $^{+ 0.54 }_{ -0.43 }$ & >120 & 0.85 & 0.05 & 0.46 & 0.03 & 414 / 414 & 0 \\
134 & 1.49 $^{+0.11 }_{ -0.11 }$ & 0.60 $^{+ 0.34 }_{ -0.33 }$ & 49 $^{+ 33 }_{ -15 }$ & 1.2 & 0.52 & 0.83 & 0.36 & 385 / 380 & 1 \\
147 & 1.7 $^{+0.05 }_{ -0.05 }$ & 0.74 $^{+ 0.2 }_{ -0.18 }$ & 144 $^{+ 113 }_{ -46 }$ & 2.53 & 1.14 & 2.08 & 0.94 & 667 / 679 & 1 \\
175$^{\star}$ & 1.77 $^{+0.19 }_{ -0.21 }$ & 0.72 $^{+ 0.73 }_{ -0.58 }$ & >52 & 0.3 & 0.24 & 0.2 & 0.16 & 134 / 134 & 0 \\
191 & 1.82 $^{+0.71 }_{ 0.41 }$ & <0.48 & >16  & 0.05 & 0.11 & 0.02 & 0.05 & 59 / 75 & 0 \\
214 & 1.66 $^{+0.03 }_{ -0.03 }$ & <0.2 & 148 $^{+ 102 }_{ -43 }$ & 3.14 & 1.72 & 2.09 & 1.15 & 958 / 948 & 1 \\
216 & 1.76 $^{+0.03 }_{ -0.03 }$ & 0.75 $^{+ 0.13 }_{ -0.13 }$ & 336 $^{+ 646 }_{ -140 }$ & 1.84 & 0.01 & 1.54 & 0.01 & 991 / 961 & 1 \\
220 & 1.64 $^{+0.03 }_{ -0.03 }$ & 0.26 $^{+ 0.07 }_{ -0.07 }$ & 52 $^{+ 4 }_{ -4 }$ & 1.56 & 4.2 & 0.92 & 2.59 & 1121 / 1138 & 1 \\
224 & 1.70 $^{+0.34 }_{ -0.32 }$ & 2.47 $^{+ 2.13 }_{ -1.22 }$ & 48 $^{+ 221 }_{ -24 }$ & 0.15 & 0.18 & 0.13 & 0.17 & 83 / 88 & 1 \\
229 & 2.10 $^{+0.11 }_{ -0.12 }$ & 0.88 $^{+ 0.75 }_{ -0.32 }$ & >142 & 0.49 & 0.33 & 0.26 & 0.18 & 197 / 177 & 0 \\
234 & 1.77 $^{+0.13 }_{ -0.13 }$ & 0.47 $^{+ 0.38 }_{ -0.32 }$ & 64 $^{+ 97 }_{ -26 }$ & 0.74 & 1.27 & 0.4 & 0.7 & 244 / 248 & 1 \\
266 & 1.97 $^{+0.03 }_{ -0.03 }$ & 0.78 $^{+ 0.11 }_{ -0.11 }$ & 233 $^{+ 147 }_{ -67 }$ & 4.45 & 1.07 & 2.43 & 0.58 & 1312 / 1105 & 1 \\
269 & 1.58 $^{+0.04 }_{ -0.04 }$ & 0.95 $^{+ 0.15 }_{ -0.14 }$ & 77 $^{+ 18 }_{ -13 }$ & 1.16 & 0.04 & 1.08 & 0.04 & 956 / 962 & 1 \\
310 & 1.82 $^{+0.02 }_{ -0.02 }$ & 0.53 $^{+ 0.06 }_{ -0.06 }$ & 140 $^{+ 29 }_{ -21 }$ & 5.83 & 0.55 & 3.57 & 0.34 & 1588 / 1438 & 1 \\
314 & 2.19 $^{+0.07 }_{ -0.06 }$ & <0.39 & >88 & 1.04 & 5.4 & 0.32 & 1.67 & 232 / 293 & 0 \\
328 & 1.77 $^{+0.11 }_{ -0.06 }$ & <0.35 & >109 & 0.25 & 0.45 & 0.16 & 0.3 & 198 / 230 & 0 \\
376$^{\star}$ & 1.63 $^{+0.21 }_{ -0.23 }$ & <0.72 & >38 & 0.24 & 1.22 & 0.15 & 0.82 & 107 / 113 & 0 \\
378 & 1.81 $^{+ 0.12 }_{ -0.12 }$ & 1.05 $^{+ 0.72 }_{ -0.28 }$ & >257  & 0.29 & 0.11 & 0.3 & 0.11 & 166 / 153 & 0 \\
402 & 1.94 $^{+ 0.06 }_{ -0.08 }$ & 0.56 $^{+ 0.31 }_{ -0.25 }$ & >159 & 1.3 & 0.5 & 0.74 & 0.28 & 420 / 427 & 0 \\
409 & 1.92 $^{+ 0.1 }_{ -0.11 }$ & 0.88 $^{+ 0.56 }_{ -0.45 }$ & >182 & 0.7 & 1.77 & 0.53 & 1.33 & 176 / 221 & 0 \\
423 & 1.90 $^{+ 0.08 }_{ -0.08 }$ & 0.94 $^{+ 0.34 }_{ -0.31 }$ & 127 $^{+ 246 }_{ -54 }$ & 1.48 & 0.34 & 0.87 & 0.2 & 465 / 439 & 1 \\
425$^{\star}$ & 1.78 $^{+ 0.11 }_{ -0.11 }$ & 0.31 $^{+ 0.37 }_{ -0.28 }$ & >79 & 1.11 & 12.04 & 0.67 & 7.53 & 285 / 322 & 0 \\
431 & 1.47 $^{+ 0.18 }_{ -0.03 }$ & <0.51 & 102 $^{+ 181 }_{ -67 }$ & 0.49 & 2.39 & 0.37 & 1.77 & 93 / 110 & 1 \\
447 & 1.81 $^{+ 0.03 }_{ -0.03 }$ & 0.85 $^{+ 0.12 }_{ -0.12 }$ & 96 $^{+ 24 }_{ -16 }$ & 1.5 & 1.17 & 1.0 & 0.79 & 1163 / 994 & 1 \\
455 & 1.92 $^{+ 0.06 }_{ -0.06 }$ & 0.53 $^{+ 0.24 }_{ -0.22 }$ & >153 & 2.36 & 0.57 & 1.35 & 0.32 & 495 / 535 & 0 \\
458 & 1.74 $^{+ 0.02 }_{ -0.02 }$ & 0.17 $^{+ 0.04 }_{ -0.04 }$ & 93 $^{+ 13 }_{ -10 }$ & 4.02 & 1.15 & 2.27 & 0.66 & 1619 / 1539 & 1 \\
473 & 1.57 $^{+ 0.10 }_{ -0.04 }$ & <0.25 & 74 $^{+ 22 }_{ -29 }$ & 1.37 & 2.49 & 0.8 & 1.46 & 302 / 315 & 1 \\
485 & 1.49 $^{+ 0.30 }_{ 0.08 }$ & <0.47 & >21 & 0.18 & 0.02 & 0.12 & 0.01 & 97 / 83 & 0 \\
495 & 1.52 $^{+ 0.70 }_{ 0.25 }$ & <0.85 & >11 & 0.06 & 0.05 & 0.03 & 0.02 & 53 / 41 & 0 \\
497 & 1.62 $^{+ 0.03 }_{ -0.03 }$ & 0.71 $^{+ 0.10 }_{ -0.10 }$ & 94 $^{+ 19 }_{ -14 }$ & 3.91 & 0.01 & 3.41 & 0.01 & 1186 / 1112 & 1 \\
512 & 1.71 $^{+ 0.11 }_{ -0.12 }$ & 0.34 $^{+ 0.36 }_{ -0.31 }$ & 90 $^{+ 640 }_{ -44 }$ & 0.54 & 0.29 & 0.34 & 0.18 & 282 / 275 & 1 \\
524 & 1.53 $^{+ 0.15 }_{ -0.16 }$ & 0.49 $^{+ 0.48 }_{ -0.4 }$ & 58 $^{+ 130 }_{ -26 }$ & 0.44 & 0.13 & 0.32 & 0.09 & 175 / 187 & 1 \\
530 & 1.60 $^{+ 0.07 }_{ -0.07 }$ & 1.29 $^{+ 0.29 }_{ -0.26 }$ & 89 $^{+ 48 }_{ -24 }$ & 0.71 & 0.01 & 0.65 & 0.01 & 684 / 656 & 1 \\
532 & 1.80 $^{+ 0.07 }_{ -0.09 }$ & 0.32 $^{+ 0.31 }_{ -0.25 }$ & >105  & 0.85 & 0.16 & 0.58 & 0.11 & 342 / 371 & 0 \\
542 & 1.74 $^{+ 0.19 }_{ -0.21 }$ & 1.01 $^{+ 0.98 }_{ -0.72 }$ & >48 & 0.24 & 0.02 & 0.21 & 0.02 & 128 / 120 & 0 \\
552 & 2.03 $^{+ 0.12 }_{ -0.11 }$ & 1.03 $^{+ 0.74 }_{ -0.62 }$ & >270  & 0.45 & 0.09 & 0.29 & 0.06 & 168 / 159 & 0 \\
556 & 1.71 $^{+ 0.09 }_{ -0.09 }$ & 0.47 $^{+ 0.29 }_{ -0.26 }$ & 75 $^{+ 68 }_{ -25 }$ & 1.04 & 0.89 & 0.68 & 0.59 & 371 / 405 & 1 \\
558 & 1.56 $^{+ 0.03 }_{ -0.03 }$ & 1.05 $^{+ 0.13 }_{ -0.12 }$ & 76 $^{+ 12 }_{ -10 }$ & 3.91 & 0.08 & 3.74 & 0.08 & 1148 / 1088 & 1 \\
565 & 1.75 $^{+ 0.1 }_{ -0.10 }$ & 0.34 $^{+ 0.34 }_{ -0.29 }$ & >79 & 0.86 & 0.29 & 0.6 & 0.2 & 293 / 300 & 0 \\
566 & 1.70 $^{+ 0.04 }_{ -0.04 }$ & 0.48 $^{+ 0.14 }_{ -0.14 }$ & 152 $^{+ 131 }_{ -50 }$ & 1.37 & 0.01 & 1.09 & 0.01 & 765 / 786 & 1 \\
567 & 1.81 $^{+ 0.05 }_{ -0.05 }$ & 0.46 $^{+ 0.17 }_{ -0.16 }$ & 92 $^{+ 48 }_{ -25 }$ & 2.73 & 0.68 & 1.66 & 0.42 & 586 / 646 & 1 \\
574 & 1.98 $^{+ 0.14 }_{ -0.14 }$ & 1.17 $^{+ 0.66 }_{ -0.51 }$ & >72 & 0.36 & 0.63 & 0.23 & 0.41 & 275 / 280 & 0 \\
576$^{\star}$ & 1.65 $^{+ 0.36 }_{ -0.37 }$ & 1.49 $^{+ 1.77 }_{ -1.08 }$ & >27 & 0.19 & 0.11 & 0.14 & 0.08 & 49 / 66 & 0 \\
583 & 1.76 $^{+ 0.09 }_{ -0.10 }$ & 0.33 $^{+ 0.32 }_{ -0.28 }$ & >81 & 0.9 & 0.08 & 0.6 & 0.05 & 281 / 313 & 0 \\
585 & 1.92 $^{+ 0.02 }_{ -0.02 }$ & 1.1 $^{+ 0.08 }_{ -0.07 }$ & >846 & 1.71 & 0.0 & 1.37 & 0.0 & 1677 / 1577 & 0 \\
587 & 1.61 $^{+ 0.14 }_{ -0.15 }$ & <0.62 & 73 $^{+ 319 }_{ -35 }$ & 0.38 & 2.79 & 0.25 & 1.87 & 167 / 214 & 1 \\
589 & 1.86 $^{+ 0.19 }_{ -0.25 }$ & 1.13 $^{+ 1.14 }_{ -0.81 }$ & >54 & 0.17 & 0.16 & 0.14 & 0.14 & 79 / 88 & 0 \\
608 & 2.09 $^{+ 0.03 }_{ -0.03 }$ & 0.71 $^{+ 0.13 }_{ -0.12 }$ & 242 $^{+ 292 }_{ -88 }$ & 2.4 & 0.09 & 1.01 & 0.04 & 1054 / 925 & 1 \\
611 & 1.92 $^{+ 0.08 }_{ -0.11 }$ & 0.37 $^{+ 0.39 }_{ -0.3 }$ & >108 & 0.92 & 1.12 & 0.51 & 0.62 & 268 / 289 & 0 \\
623 & 2.07 $^{+ 0.15 }_{ -0.17 }$ & 1.29 $^{+ 1.03 }_{ -0.68 }$ & >81 & 0.42 & 0.4 & 0.24 & 0.23 & 147 / 158 & 0 \\
631$^{\star}$ & 1.85 $^{+ 0.05 }_{ -0.05 }$ & 0.79 $^{+ 0.22 }_{ -0.2 }$ & >218 & 2.47 & 0.04 & 1.81 & 0.03 & 637 / 644 & 0 \\
636 & 1.77 $^{+ 0.04 }_{ -0.06 }$ & 0.19 $^{+ 0.17 }_{ -0.14 }$ & >282 & 3.12 & 0.42 & 2.1 & 0.28 & 329 / 304 & 0 \\
644 & 1.55 $^{+ 0.13 }_{ -0.09 }$ & <0.39 & 91 $^{+ 100 }_{ -50 }$ & 0.61 & 0.5 & 0.42 & 0.34 & 182 / 221 & 1 \\
686 & 1.59 $^{+ 0.07 }_{ -0.07 }$ & 0.74 $^{+ 0.24 }_{ -0.22 }$ & 115 $^{+ 91 }_{ -37 }$ & 2.0 & 0.01 & 1.82 & 0.0 & 631 / 569 & 1 \\
690 & 2.00 $^{+ 0.22 }_{ -0.24 }$ & 1.13 $^{+ 1.55 }_{ -0.85 }$ & >70  & 0.3 & 7.7 & 0.2 & 5.41 & 100 / 107 & 0 \\
694 & 1.71 $^{+ 0.01 }_{ -0.01 }$ & 0.43 $^{+ 0.03 }_{ -0.03 }$ & 132 $^{+ 12 }_{ -10 }$ & 11.3 & 0.65 & 8.42 & 0.48 & 2267 / 2014 & 1 \\
695 & 1.61 $^{+ 0.12 }_{ -0.12 }$ & 0.53 $^{+ 0.38 }_{ -0.32 }$ & 88 $^{+ 187 }_{ -38 }$ & 0.75 & 0.18 & 0.59 & 0.15 & 235 / 243 & 1 \\
697 & 1.97 $^{+ 0.05 }_{ -0.05 }$ & 0.54 $^{+ 0.21 }_{ -0.19 }$ & >163 & 2.66 & 0.56 & 1.41 & 0.3 & 700 / 643 & 0 \\
717 & 1.54 $^{+ 0.03 }_{ -0.03 }$ & 0.59 $^{+ 0.09 }_{ -0.09 }$ & 63 $^{+ 9 }_{ -7 }$ & 4.22 & 0.28 & 3.36 & 0.22 & 1193 / 1091 & 1 \\
726 & 2.01 $^{+ 0.08 }_{ -0.09 }$ & 0.77 $^{+ 0.41 }_{ -0.34 }$ & >215 & 0.81 & 2.54 & 0.48 & 1.51 & 328 / 326 & 0 \\
728 & 2.06 $^{+ 0.09 }_{ -0.09 }$ & 1.28 $^{+ 0.48 }_{ -0.39 }$ & >151 & 0.85 & 1.58 & 0.53 & 0.98 & 369 / 419 & 0 \\
730 & 2.15 $^{+ 0.16 }_{ -0.19 }$ & 1.52 $^{+ 1.29 }_{ -0.81 }$ & >73 & 0.36 & 0.18 & 0.2 & 0.1 & 131 / 135 & 0 \\
735$^{\star}$ & 1.73 $^{+ 0.10 }_{ -0.10 }$ & 0.79 $^{+ 0.41 }_{ -0.34 }$ & >107 & 0.88 & 0.2 & 0.66 & 0.15 & 315 / 304 & 0 \\
741$^{\star}$ & 1.91 $^{+ 0.21 }_{ -0.21 }$ & 1.11 $^{+ 0.97 }_{ -0.69 }$ & >58 & 0.46 & 0.31 & 0.29 & 0.19 & 143 / 123 & 0 \\
750 & 1.93 $^{+ 0.05 }_{ -0.05 }$ & 0.67 $^{+ 0.17 }_{ -0.16 }$ & 258 $^{+ 1078 }_{ -120 }$ & 0.94 & 0.05 & 0.55 & 0.03 & 753 / 768 & 1 \\
753 & 1.77 $^{+ 0.07 }_{ -0.07 }$ & 0.52 $^{+ 0.23 }_{ -0.21 }$ & 107 $^{+ 123 }_{ -39 }$ & 1.85 & 0.57 & 1.11 & 0.34 & 510 / 539 & 1 \\
754 & 1.96 $^{+ 0.05 }_{ -0.13 }$ & 1.18 $^{+ 0.67 }_{ -0.51 }$ & >118 & 0.53 & 0.16 & 0.4 & 0.12 & 189 / 211 & 0 \\
774 & 1.57 $^{+ 0.09 }_{ -0.09 }$ & 0.46 $^{+ 0.37 }_{ -0.26 }$ & 96 $^{+ 152 }_{ -38 }$ & 0.81 & 0.16 & 0.67 & 0.13 & 320 / 353 & 1 \\
794 & 1.77 $^{+ 0.21 }_{ -0.21 }$ & 1.29 $^{+ 0.94 }_{ -0.69 }$ & 53 $^{+ 118 }_{ -24 }$ & 0.37 & 3.51 & 0.21 & 2.03 & 152 / 164 & 1 \\
795 & 1.69 $^{+ 0.09 }_{ -0.09 }$ & 0.92 $^{+ 0.37 }_{ -0.33 }$ & 93 $^{+ 119 }_{ -36 }$ & 1.1 & 0.06 & 0.83 & 0.05 & 402 / 387 & 1 \\
810 & 2.04 $^{+ 0.10 }_{ -0.11 }$ & 1.22 $^{+ 0.45 }_{ -0.48 }$ & >171 & 0.56 & 0.54 & 0.36 & 0.35 & 238 / 252 & 0 \\
815 & 1.91 $^{+ 0.12 }_{ -0.15 }$ & 0.62 $^{+ 0.61 }_{ -0.45 }$ & >86  & 0.33 & 0.05 & 0.2 & 0.03 & 192 / 177 & 0 \\
833$^{\star}$ & 1.79 $^{+ 0.27 }_{ -0.29 }$ & 0.89 $^{+ 1.04 }_{ -0.78 }$ & >28 & 0.22 & 0.05 & 0.13 & 0.03 & 96 / 109 & 0 \\
888 & 2.01 $^{+ 0.17 }_{ -0.20 }$ & 0.61 $^{+ 0.96 }_{ -0.44 }$ & >60  & 0.23 & 0.38 & 0.13 & 0.22 & 94 / 110 & 0 \\
905 & 1.90 $^{+ 0.09 }_{ -0.09 }$ & 0.87 $^{+ 0.36 }_{ -0.31 }$ & >120  & 1.16 & 0.37 & 0.82 & 0.26 & 454 / 408 & 0 \\
907 & 1.57 $^{+ 0.17 }_{ -0.09 }$ & <0.54 & >44  & 0.28 & 2.62 & 0.24 & 2.21 & 158 / 158 & 0 \\
924 & 1.79 $^{+ 0.38 }_{ -0.12 }$ & <1.21 & >12  & 0.09 & 0.07 & 0.04 & 0.03 & 45 / 52 & 0 \\
925 & 2.08 $^{+ 0.20 }_{ -0.24 }$ & 0.91 $^{+ 1.25 }_{ -0.77 }$ & >47  & 0.28 & 0.27 & 0.15 & 0.14 & 86 / 94 & 0 \\
948 & 1.75 $^{+ 0.07 }_{ -0.07 }$ & 0.68 $^{+ 0.24 }_{ -0.22 }$ & 66 $^{+ 36 }_{ -18 }$ & 2.44 & 0.73 & 1.37 & 0.41 & 560 / 553 & 1 \\
967 & 1.83 $^{+ 0.09 }_{ -0.09 }$ & 0.28 $^{+ 0.27 }_{ -0.23 }$ & 114 $^{+ 159 }_{ -44 }$ & 1.61 & 44.73 & 0.76 & 22.07 & 418 / 460 & 1 \\
984 & 1.65 $^{+ 0.03 }_{ -0.03 }$ & 0.18 $^{+ 0.08 }_{ -0.07 }$ & 96 $^{+ 24 }_{ -16 }$ & 3.34 & 2.66 & 2.05 & 1.63 & 1151 / 1116 & 1 \\
994 & 1.66 $^{+ 0.03 }_{ -0.03 }$ & 0.22 $^{+ 0.09 }_{ -0.08 }$ & 94 $^{+ 25 }_{ -16 }$ & 4.39 & 3.26 & 2.81 & 2.1 & 1071 / 1031 & 1 \\
1013 & 1.45 $^{+ 0.11 }_{ -0.11 }$ & 0.63 $^{+ 0.34 }_{ -0.29 }$ & 43 $^{+ 20 }_{ -11 }$ & 0.82 & 1.39 & 0.64 & 1.12 & 304 / 310 & 1 \\
1021 & 1.74 $^{+ 0.15 }_{ -0.15 }$ & 1.05 $^{+ 0.59 }_{ -0.48 }$ & 54 $^{+ 76 }_{ -22 }$ & 0.53 & 0.09 & 0.32 & 0.06 & 220 / 202 & 1 \\
1032 & 1.86 $^{+ 0.03 }_{ -0.03 }$ & 0.56 $^{+ 0.10 }_{ -0.09 }$ & 201 $^{+ 110 }_{ -54 }$ & 2.65 & 0.79 & 1.65 & 0.49 & 1173 / 1104 & 1 \\
1041 & 1.84 $^{+ 0.11 }_{ -0.11 }$ & 0.81 $^{+ 0.38 }_{ -0.32 }$ & 68 $^{+ 60 }_{ -23 }$ & 2.17 & 0.04 & 0.75 & 0.02 & 838 / 779 & 1 \\
1042 & 2.19 $^{+ 0.04 }_{ -0.04 }$ & 0.91 $^{+ 0.17 }_{ -0.16 }$ & 70 $^{+ 22 }_{ -14 }$ & 2.17 & 0.04 & 0.75 & 0.02 & 838 / 779 & 1 \\
1043 & 1.80 $^{+ 0.11 }_{ -0.11 }$ & 0.63 $^{+ 0.39 }_{ -0.34 }$ & 92 $^{+ 244 }_{ -41 }$ & 0.9 & 0.34 & 0.52 & 0.19 & 290 / 305 & 1 \\
1046 & 1.72 $^{+ 0.02 }_{ -0.02 }$ & 0.56 $^{+ 0.06 }_{ -0.06 }$ & 99 $^{+ 15 }_{ -11 }$ & 3.78 & 0.02 & 2.53 & 0.02 & 1522 / 1475 & 1 \\
1082 & 1.63 $^{+ 0.10 }_{ -0.06 }$ & <0.28 & 91 $^{+ 48 }_{ -40 }$ & 0.96 & 8.05 & 0.51 & 4.23 & 284 / 297 & 1 \\
1084$^{\star}$ & 1.54 $^{+ 0.21 }_{ -0.23 }$ & 0.36 $^{+ 0.67 }_{ -0.34 }$ & >36 & 0.35 & 0.18 & 0.28 & 0.15 & 99 / 98 & 0 \\
1088 & 1.75 $^{+ 0.03 }_{ -0.03 }$ & 0.52 $^{+ 0.09 }_{ -0.08 }$ & 86 $^{+ 15 }_{ -11 }$ & 2.96 & 8.06 & 1.82 & 5.05 & 1115 / 1126 & 1 \\
1090 & 1.71 $^{+ 0.01 }_{ -0.01 }$ & 0.38 $^{+ 0.04 }_{ -0.04 }$ & 75 $^{+ 7 }_{ -6 }$ & 4.74 & 1.29 & 2.98 & 0.82 & 1746 / 1591 & 1 \\
1106 & 1.82 $^{+ 0.11 }_{ -0.11 }$ & 0.63 $^{+ 0.38 }_{ -0.33 }$ & 89 $^{+ 199 }_{ -38 }$ & 1.05 & 0.64 & 0.57 & 0.35 & 313 / 326 & 1 \\
1111 & 1.81 $^{+ 0.03 }_{ -0.03 }$ & 0.74 $^{+ 0.10 }_{ -0.09 }$ & 72 $^{+ 11 }_{ -9 }$ & 3.16 & 0.15 & 1.93 & 0.09 & 1015 / 1075 & 1 \\
1117 & 1.67 $^{+ 0.11 }_{ -0.11 }$ & 0.61 $^{+ 0.36 }_{ -0.31 }$ & 55 $^{+ 50 }_{ -19 }$ & 1.03 & 0.15 & 0.63 & 0.09 & 305 / 289 & 1 \\
1125 & 1.89 $^{+ 0.22 }_{ -0.24 }$ & 0.72 $^{+ 1.12 }_{ -0.54 }$ & >40  & 0.24 & 3.05 & 0.15 & 1.89 & 107 / 108 & 0 \\
1162 & 1.58 $^{+ 0.14 }_{ -0.14 }$ & 0.47 $^{+ 0.44 }_{ -0.36 }$ & 72 $^{+ 186 }_{ -33 }$ & 0.44 & 0.06 & 0.34 & 0.05 & 191 / 187 & 1 \\
1172 & 1.62 $^{+ 0.04 }_{ -0.03 }$ & <0.09 & 77 $^{+ 19 }_{ -14 }$ & 5.95 & 5.83 & 3.38 & 3.33 & 831 / 856 & 1 \\
1182 & 1.92 $^{+ 0.05 }_{ -0.05 }$ & 0.67 $^{+ 0.19 }_{ -0.18 }$ & >244  & 2.98 & 0.18 & 1.97 & 0.12 & 686 / 688 & 0 \\
1381 & 1.84 $^{+ 0.22 }_{ -0.22 }$ & 0.89 $^{+ 0.95 }_{ -0.71 }$ & 56 $^{+ 261 }_{ -27 }$ & 0.42 & 12.6 & 0.19 & 6.6 & 163 / 148 & 1 \\
1471$^{\star}$ & 1.72 $^{+ 0.13 }_{ -0.13 }$ & 0.88 $^{+ 0.54 }_{ -0.44 }$ & >81 & 0.43 & 0.11 & 0.36 & 0.09 & 197 / 223 & 0 \\
1503$^{\star}$ & 1.63 $^{+ 0.16 }_{ -0.16 }$ & 0.65 $^{+ 0.58 }_{ -0.47 }$ & >56 & 0.36 & 0.15 & 0.3 & 0.13 & 147 / 161 & 0 \\
1536 & 1.47 $^{+ 0.12 }_{ -0.12 }$ & 0.61 $^{+ 0.40 }_{ -0.33 }$ & 60 $^{+ 49 }_{ -20 }$ & 0.63 & 2.42 & 0.53 & 2.07 & 265 / 280 & 1 \\
1581 & 1.73 $^{+ 0.23 }_{ -0.23 }$ & 0.66 $^{+ 0.96 }_{ -0.46 }$ & >40  & 0.32 & 5.95 & 0.23 & 4.18 & 105 / 110 & 0 \\

\hline
\end{longtable}
\noindent{$^{a}$ Un-absorbed flux in units of \funits \\
$^{b}$ Un-absorbed luminosity in units of \lunits \\
$^{c}$ Identifier for the high energy cut-off confidence interval estimation. A value of 0 denotes lower limit measurement, a value of 1 corresponds to a securely inferred value}
}

\begin{table*}
\centering
\caption{Mean values for the spectral parameters}
\begin{tabular}{ccc}
\hline
              &  Direct estimates  &  Survival analysis  \\ 
              \hline
 Photon Index   & 1.78$\pm0.01$    (118/118)  &         -   \\
 Reflection     & 0.68$\pm0.05$    (106/118)  & $0.69\pm0.04$    \\
 Cut-off energy & 102$\pm16$ keV    (63/118)  & $206\pm38$     \\
 \hline
 \multicolumn{3}{c}{\footnotesize Notes: As the cut-off energy distribution is highly skewed, often presenting large asymmetric error-bars the error approximation is unrealistic.} 
\end{tabular}
    \label{tab:summary}
\end{table*}

\bibliography{ref}{}
\bibliographystyle{aa}
\end{document}